%% file: down-going_nus_arxiv.tex
\documentclass[aps,prd,twocolumn,showpacs,superscriptaddress, altaffilletter]{revtex4-1}  

\usepackage{graphicx}  
\usepackage{dcolumn}   
\usepackage{bm}        
\usepackage{amssymb}   

\usepackage{amsthm}    

\usepackage{xspace}
\def\Offline{\mbox{$\overline{\textrm%
{Off}}$\hspace{.05em}\raisebox{.4ex}{$\underline{\textrm{line}}$}}\xspace}

\begin{document}

\title{A search for ultra-high energy neutrinos in highly inclined events at the Pierre Auger Observatory}

\include{auger_authors_RevTex_noBlank}

\begin{abstract}
The Surface Detector of the Pierre Auger Observatory is sensitive to neutrinos of all flavours above 0.1 EeV. These
interact through charged and neutral currents in the atmosphere giving rise to extensive air showers. When interacting deeply in the
atmosphere at nearly horizontal incidence, neutrinos can be distinguished from regular hadronic cosmic rays by the 
broad time structure of their shower signals in the water-Cherenkov detectors.
In this paper we present for the first time an analysis based on down-going neutrinos. We describe the search procedure, 
the possible sources of background, the method to compute the exposure and the associated systematic uncertainties.
No candidate neutrinos have been found in data collected from 1 January 2004 to 31 May 2010.
Assuming an $E^{-2}$ differential energy spectrum the limit on the single flavour neutrino is $E^{2}{\rm d}N/{\rm d}E < 1.74 \times 10^{-7}~{\rm GeV~cm^{-2}~s^{-1}~sr^{-1}}$ at 90\% C.L. in the energy range $1\times 10^{17} {\rm eV}< E < 1\times 10^{20} {\rm eV}$.
\end{abstract}

\maketitle

\section{Introduction}
\label{sec:intro}

Neutrinos play a key role in the understanding of the origin of ultra high
energy cosmic rays (UHECRs). Their observation should open a new window to the
universe since they can give information on regions that are otherwise hidden
by large amounts of matter in the field of view.  Moreover,
neutrinos are not deviated by magnetic fields and would point back to their sources.

In the EeV range, neutrinos are expected to be produced in the same sources
where UHECRs are thought to be accelerated, as well as during the propagation
of UHECRs through the cosmic microwave background (CMB) radiation~\cite{nu_reviews}. The latter
are called cosmogenic neutrinos and their presence is expected if the 
UHECRs above the spectral cut-off reported in~\cite{Auger_spectrum}
contain a significant fraction of protons~\cite{nu_GZK_Berezinsky, nu_GZK_Yoshida, nu_GZK_Engel, nu_GZK_Ahlers, nu_GZK_Kotera, nu_GZK_BerezinskyNew}.

There are many current programs to search for high energy neutrinos with dedicated
experiments~\cite{IceCube,Antares,ANITA}.  Although the primary goal of the
Pierre Auger Observatory Surface (SD) and Fluorescence Detectors (FD) is to
detect UHECRs, UHE neutrinos (UHE$\nu$s) can also be identified and limits to the
diffuse flux of UHE$\nu$s in the EeV range and above have been set using earlier Auger data~\cite{PRL_nutau,Tiffenberg_icrc09,Gora_icrc09}. 
Earth-skimming $\tau$ neutrinos are expected to be observed through the detection
of showers induced by the decay of emerging $\tau$ leptons which are created 
by $\nu_\tau$ interactions in the Earth~\cite{nutau}. Using this mechanism for data collected
from 1 January 2004 until 30 April 2008, an upper limit was set:  
$E_{\nu}^{2} dN_{\nu_{\tau}}\slash dE_{\nu}< 6^{+3}_{-3}\times10^{-8}\,{\rm GeV\,cm}^{-2} {\rm s}^{-1} {\rm sr}^{-1}$ at
90\% CL for each neutrino flavour~\cite{nu_tau_long}.
The SD of the Pierre Auger Observatory has also been shown to be sensitive to 
``down-going'' neutrinos of all flavours interacting in the atmosphere or in 
the mountains surrounding the SD,
and inducing a shower close to the ground~\cite{Gora_icrc09,Idea_Detection,nu_down_Auger}.
In this paper we present an analysis based
on down-going neutrinos and place a competitive limit on the all-flavour
diffuse neutrino flux using data from 1 January 2004 until 31 May 2010. 

The main challenge in detecting UHE neutrinos with the Pierre Auger Observatory 
is to identify a neutrino-induced shower in the
background of showers initiated by UHECRs, possibly protons or heavy nuclei~\cite{Auger_Xmax_PRL}
and, in a much smaller proportion, even photons~\cite{Auger_photon_limit}.

\begin{figure*}[ht]
\begin{center}
\noindent
\includegraphics [width=17.8cm]{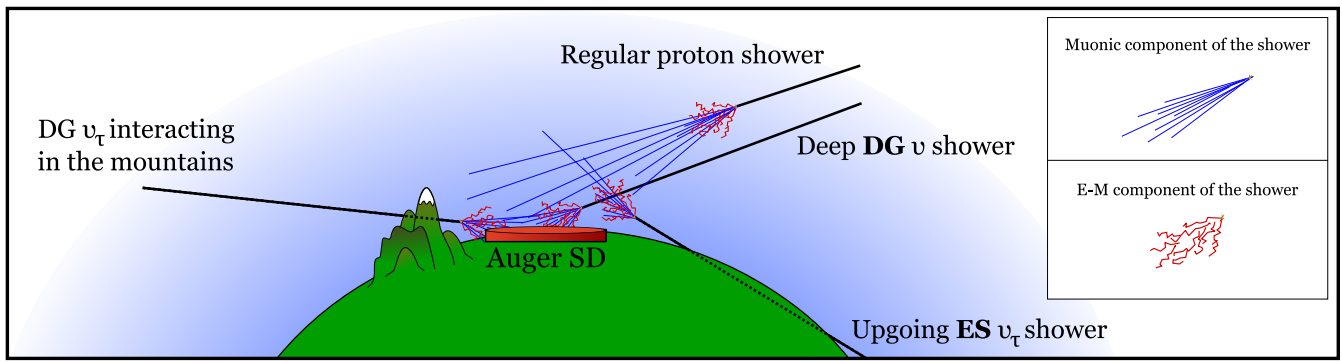}
\end{center}
\caption{Pictorial representation of 
the different types of showers induced by protons, heavy nuclei
and ``down-going" (DG) as well as ``Earth-skimming" (ES) neutrinos. The search
for down-going showers initiated deep in the atmosphere is the subject
of this work.}
\label{fig:all_neutrinos_auger}
\end{figure*}

The identification of $\nu$-induced showers is illustrated in Fig.~\ref{fig:all_neutrinos_auger}. If the
incidence is nearly horizontal, ``old'' showers induced in the upper atmosphere
by protons, nuclei or photons have a thin and flat front at ground level,
containing only high energy muons and their radiative and decay products,
concentrated within a few tens of nanoseconds. On the other hand, ``young'' showers, induced
by neutrinos at a low altitude, have a thick, curved front with a significant
electromagnetic component spread in time over hundreds of nanoseconds,
specially in their earlier part that traverses less atmosphere. In this work, to obtain an unambiguous
identification of neutrinos, we select showers with zenith angle $\theta > 75^{\circ}$ and
we apply criteria to ensure a deep interaction. Using less inclined showers is
in principle possible, but will require a better control of the various sources
of background.

The method was tuned using data taken at the SD in the period
from 1 January 2004 until 31 October 2007. A blind scan over the data collected in
the remaining period, i.e., from 1 November 2007 until 31 May 2010
reveals no candidates and we place a stringent limit on the diffuse flux of UHE neutrinos. 

For that purpose, we calculate the probability for a shower, produced deeply 
in the atmosphere, to trigger the SD and to be identified as a neutrino candidate.
This probability depends on the neutrino flavour and type of interaction 
-- charged current (CC) or neutral current (NC) -- and is also a function of 
neutrino energy $E_\nu$, incident zenith angle $\theta$, 
and atmospheric interaction depth.
From these identification probabilities we calculate the exposure 
of the SD to deep inclined neutrino showers. 
We give an estimate of the systematic uncertainties on the diffuse neutrino flux limit,
and discuss the implications of our observations for models of UHE neutrino production.

\section{The Pierre Auger Observatory}
\label{sec:pao}
The Pierre Auger Observatory is a hybrid
detector located in Malarg\"ue, Mendoza, Argentina~\cite{EApaper}.
It consists of an array of particle detectors~\cite{Auger_SD} and a set of 
fluorescence telescopes~\cite{Auger_FD} at four sites that provide a unique cross calibration capability.

The SD is spread over a surface of $\sim$3000~${\rm km^2}$ at an altitude of 
$\sim$1400~m above sea level. This corresponds to an average vertical atmospheric depth above
ground of $X_{ground} = 880~{\rm g~cm^{-2}}$. 
The slant depth $D$ is the total
grammage traversed by a shower measured from ground in the direction
of the incoming primary particle. In the flat-Earth approximation $D =
(X_{ground} - X_{int} )/ \cos \theta$, where $X_{int}$ is the interaction depth and $\theta$ the zenith
angle. For very inclined showers the curvature of the atmosphere is taken
into account.

The four fluorescence sites are located at the perimeter of the surface array viewing 
the atmosphere above it~\cite{Auger_FD}. 
In this work only data collected with the SD of the Pierre Auger Observatory 
are used to search for down-going neutrinos.

\subsection{The Surface Detector}
\label{sec:sd}

Since the beginning of its operation for physics analysis, in January 2004, the 
SD array has grown steadily and it has been recording an increasing 
amount of data. It consists of $\sim$1660 detector units (water-Cherenkov
stations) regularly spaced in a triangular grid of side 1.5 km. Each
detector unit is a cylindrical polyethylene tank of 3.6 m diameter 
and 1.2 m height containing 12,000 liters of purified water.
The top surface has
three photomultiplier tubes (PMTs) in optical contact with
the water in the tank. The PMT signals are sampled by flash 
analog digital converters (FADC) with a frequency of 40~MHz. 
Each surface detector is regularly monitored and calibrated 
in units of vertical equivalent muons (VEM) corresponding to 
the signal produced by a $\mu$ traversing the tank vertically and through its center~\cite{Auger_calibration}. 
The surface stations transmit information by radio links
to the Central Data Acquisition System (CDAS) located in
Malarg\"ue. The PMTs, local processor, GPS receiver,
and the radio system are powered by batteries regulated by solar
panels. Once installed, the local stations work continuously
without external intervention.

\subsection{The trigger}
\label{sec:trig}

A local trigger selects signals, either with a high peak value, or with
a long duration. The second condition favours stations hit in the early stage
of the shower development (moderately inclined or deeply induced showers).
The global trigger requires either 4 stations satisfying one of the conditions,
or 3 stations satisfying the second one, in a compact configuration (see~\cite{Auger_trigger}
for more details).

With the complete array, the global trigger
rate is about two events per minute, one half being
actual shower events with median energy of $3\times 10^{17}$~eV.

\section{Simulation of neutrino interactions, induced showers and the response of the Surface Detector.}
\label{sec:nuSim}
Monte Carlo simulations of neutrino-induced showers are used to establish
identification criteria and to compute the acceptance of the SD to UHE$\nu$s.
The whole simulation chain is divided in three stages:
\begin{enumerate}
\item High energy processes:
  \begin{itemize}
  \item The $\nu$-nucleon interaction is simulated with {\sc herwig}~\cite{HERWIG}.
  \item In the case of $\nu_{\tau}$ CC interactions, the $\tau$ lepton propagation is simulated with a dedicated code and its decay (when necessary) with {\sc tauola}~\cite{TAUOLA}.
  \end{itemize} 
\item The shower development in the atmosphere is processed by {\sc aires}~\cite{aires}.
\item The Surface Detector simulation is performed with the \Offline software ~\cite{Offline}.
\end{enumerate}
In the next subsections we discuss each stage in detail.

\subsection{Neutrino interaction}

\begin{figure*}[ht]
\begin{center}
\noindent
\includegraphics [width=17.8cm]{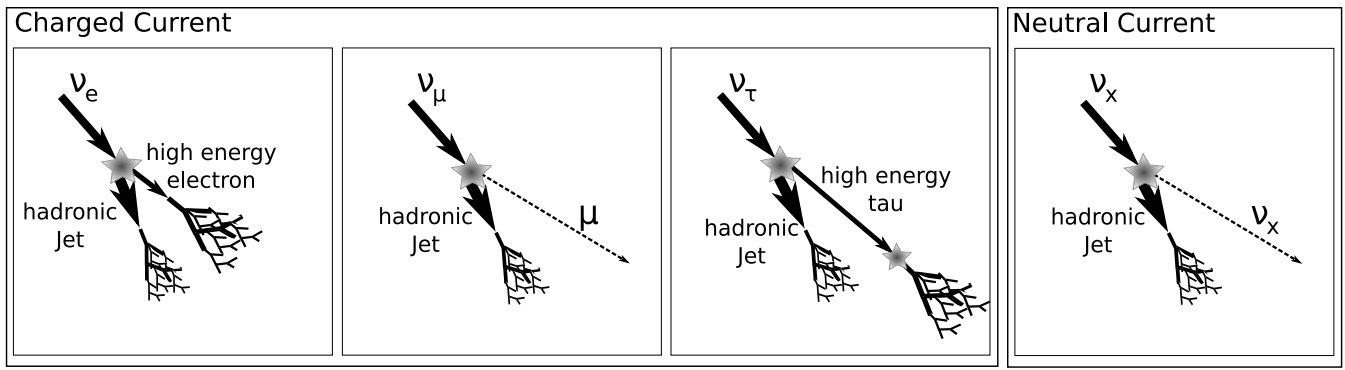}
\end{center}
\caption
{
Different types of atmospheric showers induced by neutrinos.
}
\label{fig:croquis}
\end{figure*}

{\sc herwig} is a general-purpose event generator for high-energy processes,
with particular emphasis on the detailed simulation of QCD parton showers.
Here it is used to compute the fraction of the primary energy that goes
into the hadronic vertex and to provide the secondary particles produced for
both charged~(CC) and neutral current~(NC) interactions (see Fig.~\ref{fig:croquis} for a
summary of all the channels considered in this work).

The energy carried by the hadronic jet is always converted into a shower which could be
seen by the SD. In addition, the energy of the lepton produced in a CC interaction may be totally or partially visible. An electron is promptly converted into an electromagnetic shower. A $\tau$ at EeV energies has a decay length of $\sim50$~km and may decay before reaching the ground producing a secondary shower that can be detected (so called ``double bang'' event).
On the other hand, it is very unlikely that a high energy muon will produce a detectable shower, so its interaction and/or decay are not simulated. 
For all channels and neutrino flavours a set of primary $\nu$ interactions
is constructed from a grid of incoming neutrino energies, zenith angles and
interaction depths.
In ``double bang'' events the decay products of the $\tau$ lepton are generated by {\sc tauola}.
The energies and momenta of the secondary particles are then injected into the program {\sc aires} to generate the
atmospheric cascade.

\subsection{Down-going neutrinos interacting in the mountains }
\label{sec:mountains}

In addition to the interactions in the
atmosphere, we also take into account the possibility of $\tau$ neutrino interactions
within the mountains around the Pierre Auger Observatory (mainly the
Andes located to the north-west of the array), producing a
hadronic jet and a $\tau$ lepton. The hadronic or electromagnetic showers produced
by neutrinos of any flavour are absorbed either in the rock itself, or in the few
ten kilometers of atmosphere between the mountains and the Auger array, and may be neglected.
So only showers induced by the decay of the $\tau$s may be seen. In other terms, this process is exactly equivalent to
the ``Earth-skimming'' mechanism, but it is included in this study because such showers are going downwards.

The topography surrounding the SD of the Auger Observatory is accounted for using a digital elevation map~\cite{nasa_dem}.
For the Auger site, the line of sight intercepting the mountains corresponds only to zenith angles very close
to the horizon ($\theta > 89^{\circ}$). Even though the solid angle is much smaller than for
showers with $\theta > 75^{\circ}$, this mechanism is still relevant because mountains are
much more massive. It is simulated in the same way as the ``double bang''
process, accounting in addition for energy loss of the $\tau$ lepton in the rock~\cite{tau_prop}.

\subsection{Detector simulation}

To avoid excessively long computing times {\sc aires}
uses the standard thinning procedure~\cite{thinning} consisting in
following only some branches in the tree of interactions in the
atmosphere. Weights are attributed to the surviving branches,
obtaining a representative set of particles at any stage, especially at ground level.
The first step in the detector response simulation is to regenerate a fair sample of
the particles expected in each station from the thinned output of {\sc aires}. This unthinning procedure is 
detailed in~\cite{Billoir_unthinning}.  
Each particle reaching a surface detector station is injected in the station, and the
amount of Cherenkov light produced in water calculated with {\sc geant4}~\cite{GEANT4}. 
The FADC traces of the PMT signals are simulated using the \Offline framework~\cite{Offline}. 
The total signal due to the particles entering the station, as well as several 
quantities characterizing the FADC trace which will be relevant
for neutrino identification (see below) are then calculated.   
The local and global trigger conditions 
are applied in the same way as for real data. 

\section{Inclined event selection and reconstruction}
\label{sec:eventSelection}

\begin{figure*}[ht]
\begin{center}
\noindent
\includegraphics [width=17.8cm]{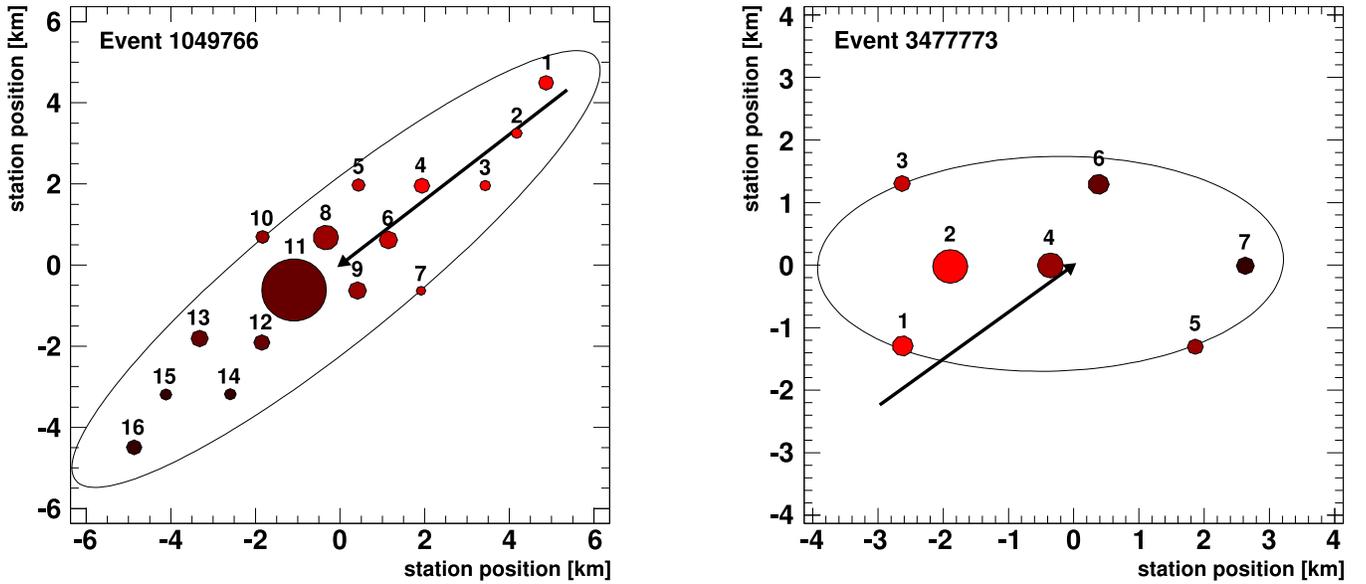}
\end{center}
\caption{\textit{Left panel:} Event produced by a nearly horizontal shower ($\theta_{rec}=80^{\circ}$). The footprint (ellipse) is elongated along the reconstructed direction of arrival (arrow).   
\textit{Right panel:} a non-inclined event with $\theta_{rec}=79^{\circ}$.  The major axis of the footprint and the reconstructed direction of arrival do not point in the same direction. Close inspection of the event suggests that stations 3 and 5 are accidental and corrupt the reconstruction.
The numbers indicate the triggering order of the stations.}
\label{fig:evts}
\end{figure*}

Events occurring during periods of data acquisition instabilities~\cite{Auger_trigger} are
excluded. After a ``trace cleaning'' procedure removing the accidental signals
(mainly atmospheric muons), the start times of the signals in the stations
are requested to be compatible with a plane shower front moving at speed $c$.
If this condition is not fulfilled
using all stations included in the global trigger, an iterative
procedure removes stations until a satisfactory configuration is found with
at least four stations. Otherwise the event is rejected.
The angle between a vertical axis and the perpendicular direction to this plane is the reconstructed zenith angle $\theta_{rec}$ of the shower.
Nearly horizontal showers are selected by requiring $\theta_{rec}>75^{\circ}$. 
In some cases a non-inclined event, produced by detector fluctuations or two independent showers arriving close in time (less that 60~ns), may be incorrectly reconstructed as inclined. To remove these events we also compute the apparent speed of propagation of the trigger between every pair of stations ($V_{ij}$) and the average speed of the event ($\langle V \rangle$), as in~\cite{nu_tau_long}.
Genuine inclined showers have a ``footprint'' (configuration of the stations) elongated in the direction of arrival (left panel of Fig.~\ref{fig:evts}).
The apparent speed of propagation of the signal, along the major axis of the footprint, is concentrated around the speed of light $c$.

\begin{figure}[ht]
\begin{center}
\noindent
\includegraphics [width=8.6cm]{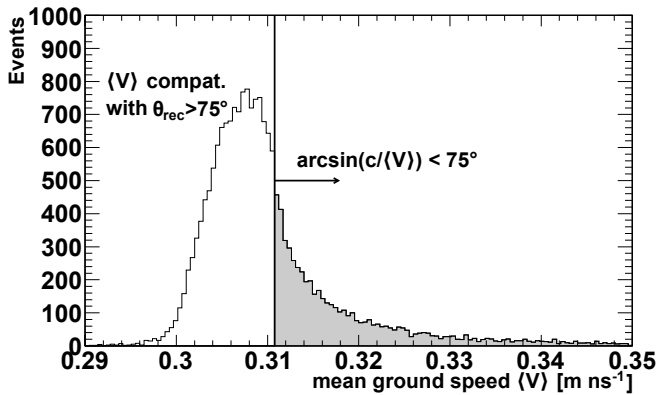}
\end{center}
\caption{Distribution of the mean ground speed of the signal for events with $\theta_{rec}>75^{\circ}$ acquired between 1 January 2004 and 31 October 2007.}
\label{fig:mean_ground_speed_RE}
\end{figure}

Under the plane front approximation, the zenith angle is $\simeq \arcsin(c/\langle V \rangle)$. In Fig.~\ref{fig:mean_ground_speed_RE} we show the distribution of $\langle V \rangle$ for events with $\theta_{rec}>75^{\circ}$ acquired between 1 January 2004 and 31 October 2007. The shaded region corresponds to misreconstructed or low quality events (see right panel of Fig.~\ref{fig:evts} for an example).
To remove these events we optimized a set of quality cuts using a MC sample of 5000 regular inclined showers initiated by hadrons near the top of the atmosphere: $\langle V \rangle$ is required to be less than 0.313~m~ns$^{-1}$, with a relative spread smaller
than 0.08\%. Also, the ``footprint'' is required to be elongated: $L/W > 3$, where $L$ and $W$ are the length and the width
(eigenvalues of the inertia tensor, as defined in~\cite{nu_tau_long}). These cuts reject only 10\% of genuine inclined showers.

For events where all stations are aligned along one of the directions of the array, $\theta_{rec}$ cannot be computed and we rely on the average speed of the event, $\langle V \rangle$. 
These ``in-line'' events are of great importance since the Monte Carlo simulations show that low energy neutrinos ($\lesssim 10^{18}$ eV) 
typically present this type of configuration in the SD.

There is an additional requirement for events constituted by an in-line event plus a non-aligned station (a non-aligned event that would become in-line by removing just one station).
This kind of spatial configuration is particularly prone to bad reconstruction if the non-aligned station was triggered by accidental muons not belonging to the shower front.
To avoid this problem we also reconstruct the in-line event obtained by the removal of the non-aligned station and require it to have mean ground speed compatible with a zenith angle larger than $75^\circ$.

\section{Identification of neutrino candidates}
\label{sec:data_and_disc}

For this analysis, the whole data period (1 Jan 04 - 31 May 10), was divided into two separate samples. 
Selected events recorded between 1 Jan 04 and 31 Oct 07  
(equivalent to $\sim1.4$ years of a complete SD array working continuosly) constitute the ``training" sample, 
used to develop and optimize the neutrino identification algorithms. 
Data collected between 1 Nov 07 and 31 May 10 (equivalent to $\sim2$ yr of the full array), 
constitute the ``search" sample. These latter events were not processed
before the final tuning of the algorithms defining the neutrino identification criteria. 

\subsection{Discrimination of neutrinos from hadronic showers}
\label{sec:fisher}
\begin{figure}[ht]
\begin{center}
\noindent
\includegraphics [width=8.6cm]{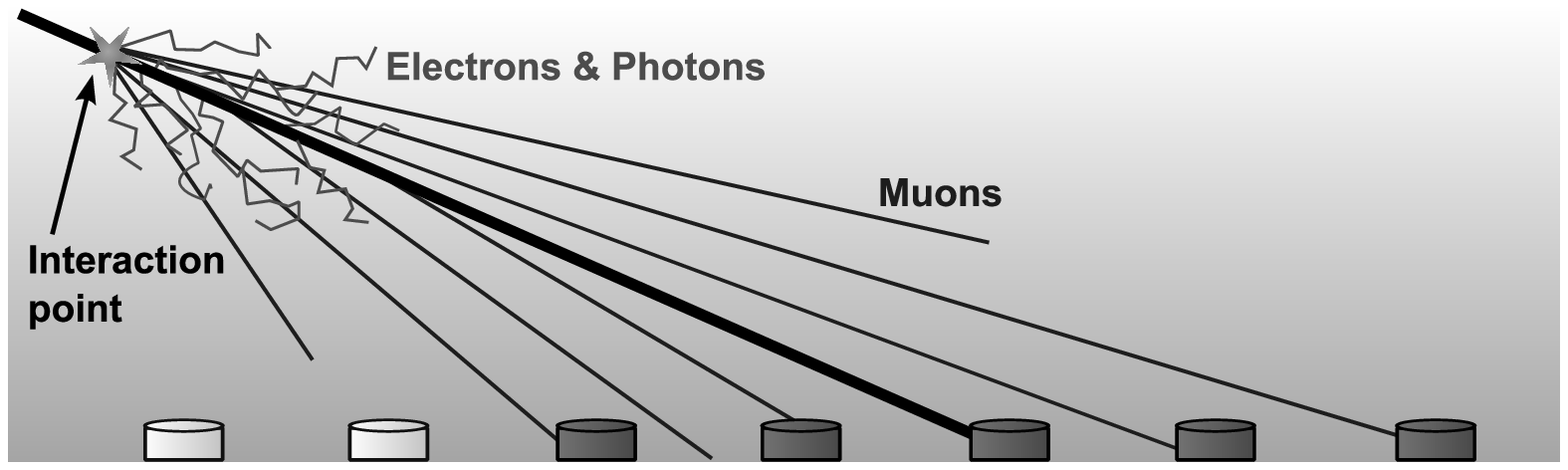} \\
\vskip +0.3cm
\includegraphics [width=8.6cm]{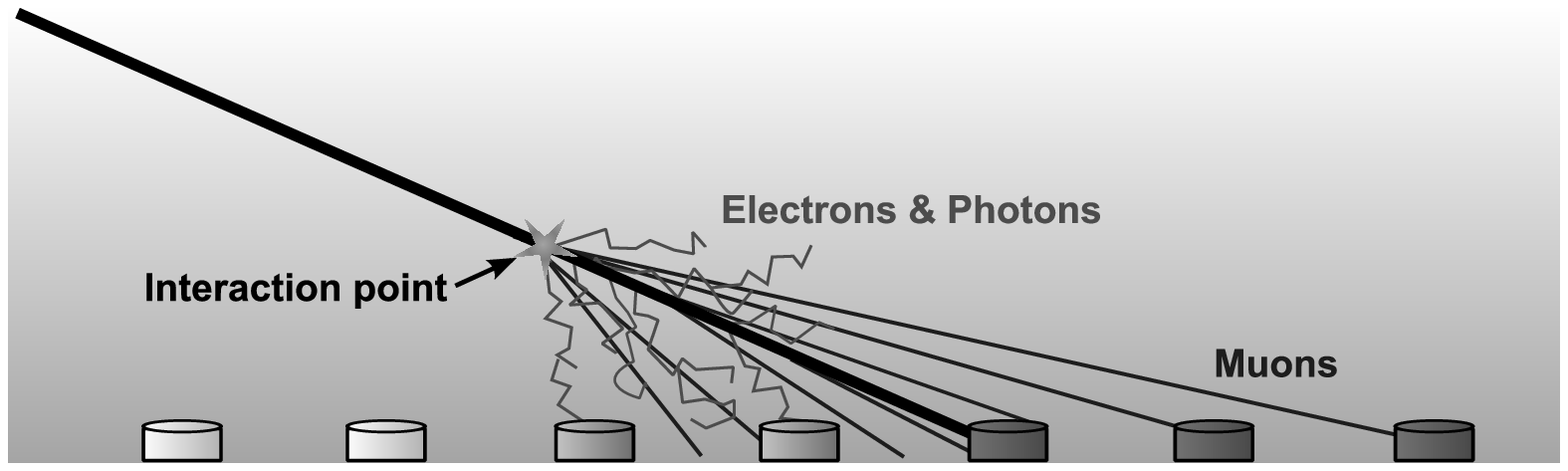}
\end{center}
\caption{\textit{Upper panel}: sketch of an inclined shower induced by a hadron interacting high in the atmosphere. The EM component is absorbed and only the muons reach the detector. \textit{Lower panel}: deep inclined shower. Its early region has a significant EM component at the detector level.}
\label{fig:shallow_deep}
\end{figure}

Neutrinos, unlike protons and heavier nuclei, can generate showers initiated
deeply into the atmosphere.  The main signature of these deep showers in the SD
is a significant electromagnetic (EM) component spread in time over hundreds of nanoseconds,
especially in the region on the ground at which the shower arrives earlier (see
Fig.~\ref{fig:shallow_deep}). On the other hand, hadron-induced showers
start high in the atmosphere, their electromagnetic component is fully
absorbed and only high energy muons and their radiative and decay products
reach the surface, concentrated within a few tens of nanoseconds.

\begin{figure*}[ht]
\begin{center}$
\begin{array}{cc}
\includegraphics [width=8.6cm]{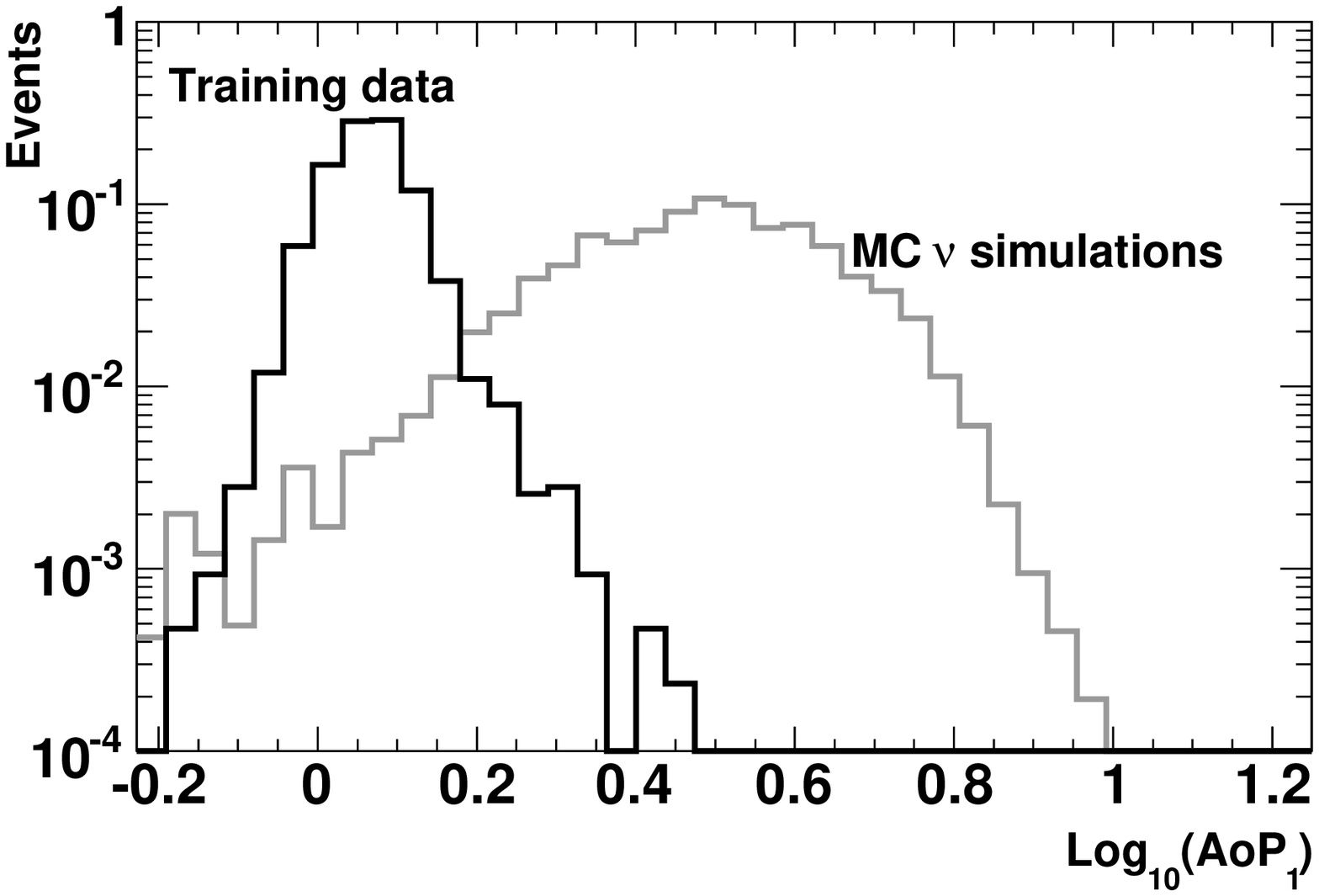} &
\includegraphics [width=8.6cm]{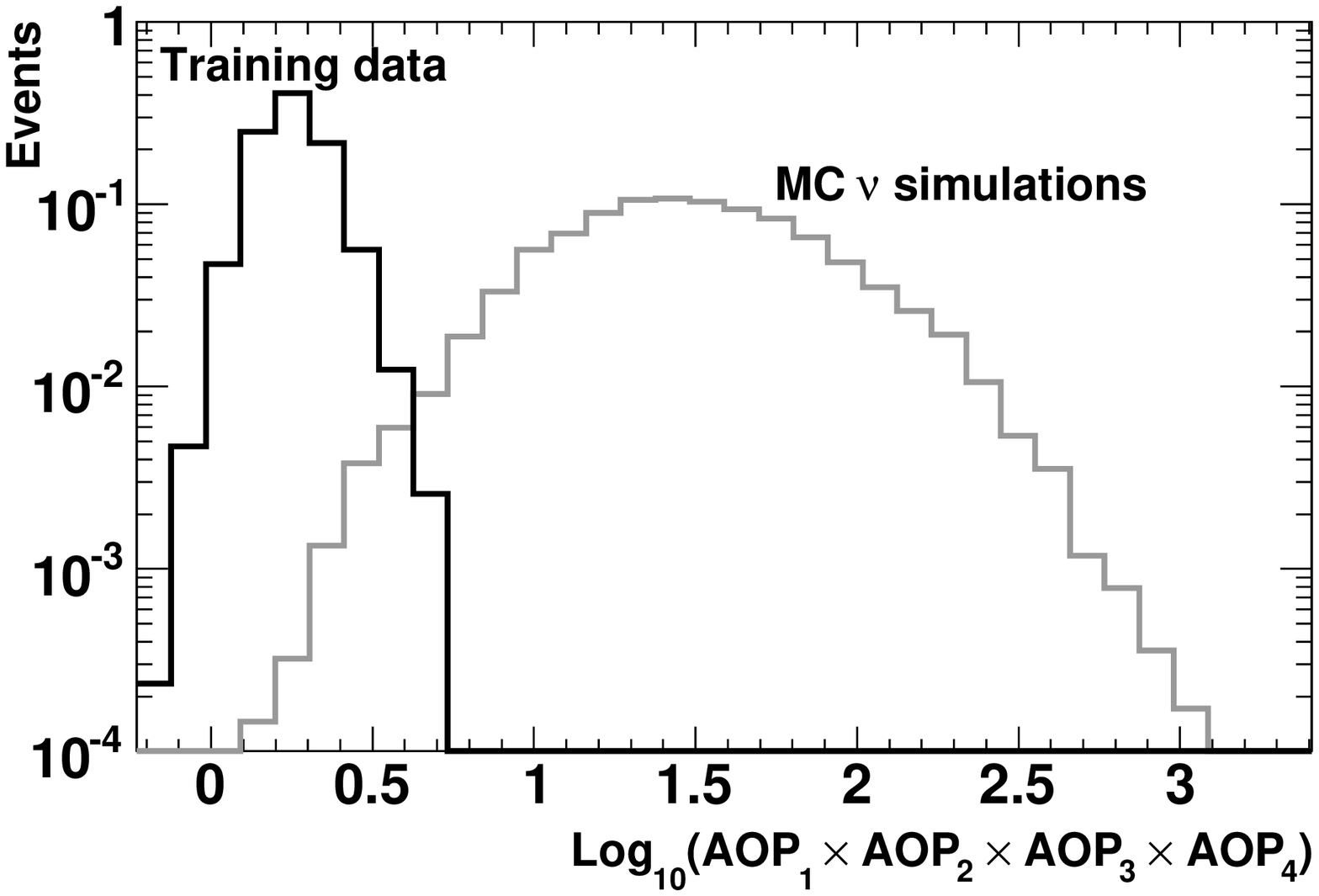}
\end{array}$
\end{center}
\caption{Distributions of the AoP of the earliest station (left) and the 
product of the first four AoP (right) in background (real events in the training sample) and
simulated $\nu_e$ CC events. There is a clear separation 
between both samples indicating that the AoP of the early stations 
is a good discrimination observable to be used in the Fisher method. See text for more details.}
\label{fig:aop}
\end{figure*}

We identify stations reached by wide EM-rich shower fronts via their Area-over-Peak ratio (AoP), 
defined as the ratio of the integral of the FADC
trace to its peak value, normalized to 1 for the average signal produced
by a single muon. In background horizontal showers the muons and their
electromagnetic products are concentrated within a short time interval, so
their AoP is close to 1. In the first stations hit by a deep inclined shower, it is typically
between 3 and 5 (see left panel of Fig.~\ref{fig:aop}).

To quantify the distinctive features of hadronic and deeply penetrating showers
induced by neutrinos at large zenith angle, improve the separation between the 
samples and enhance the efficiency, while keeping a simple physical interpretation 
of the identification process, we choose a multivariate technique known as the 
Fisher discriminant method~\cite{FDM}.
To tune it we used as a ``signal'' sample the Monte Carlo simulations -- exclusively
composed of neutrino-induced showers -- and as ``background'' the training
sample introduced above -- overwhelmingly, if not totally, constituted of
nucleonic showers.
We use real data to train the Fisher discrimination
method, instead of simulations of hadronic showers, for two main reasons:
\begin{itemize}
 \item the composition of the primary flux is not known, and moreover the
  interaction models used to simulate hadronic showers may bias some
  features of the tail of the distributions of the observables used in this
  analysis.
 \item the detector simulation may not account for all possible detector defects
  or fluctuations that may contribute to the background to ultra-high
  energy neutrinos, while the real data contain all of them, including
  those which are not well known, or even not yet diagnosed.
\end{itemize}
Note that, since we apply a statistical method for the discrimination, the use 
of real data as a background sample does not imply that we assume it contains no neutrinos, 
but just that, if any, they constitute a small fraction of the total recorded events.

After training the Fisher method, a good discrimination is found when using the following ten variables
\cite{Gora_icrc09}: the AoP of the four earliest triggered stations in
each event, their squares, their product, and a global early-late asymmetry
parameter of the event.  We include the square of the AoP because when the
distribution of the input variables is not gaussian the addition of a non-linear
combination of them improves the discrimination power~\cite{Roe_03}. The
product of the AoP of the earliest four stations in the event aims at minimizing
the relative weight of an accidentally large AoP produced, for instance, by a
single muon which does not belong to the shower front arriving at a station before or
after the shower itself. This variable is also a very good discriminator as shown
in the right panel of Fig.~\ref{fig:aop}. The early-late asymmetry parameter is a global
observable of the event defined as the difference between the mean AoP of the
earliest and latest stations in the event. We have checked in simulations that
neutrino-induced events typically have an asymmetry parameter larger than
proton or nucleus-induced showers~\cite{Gora_icrc09}.  Finally, the addition of
other observables characterizing the time spread of the signals, such as the
rise-time (between 10\% and 50\% of the integrated signal) or the fall-time
(between 50\% and 90\% ), or including local observables of the stations that
trigger last in the event, do not bring about significant improvements in the
discrimination.

\begin{figure*}
\begin{center}$
\begin{array}{cc}
\includegraphics [width=8.6cm]{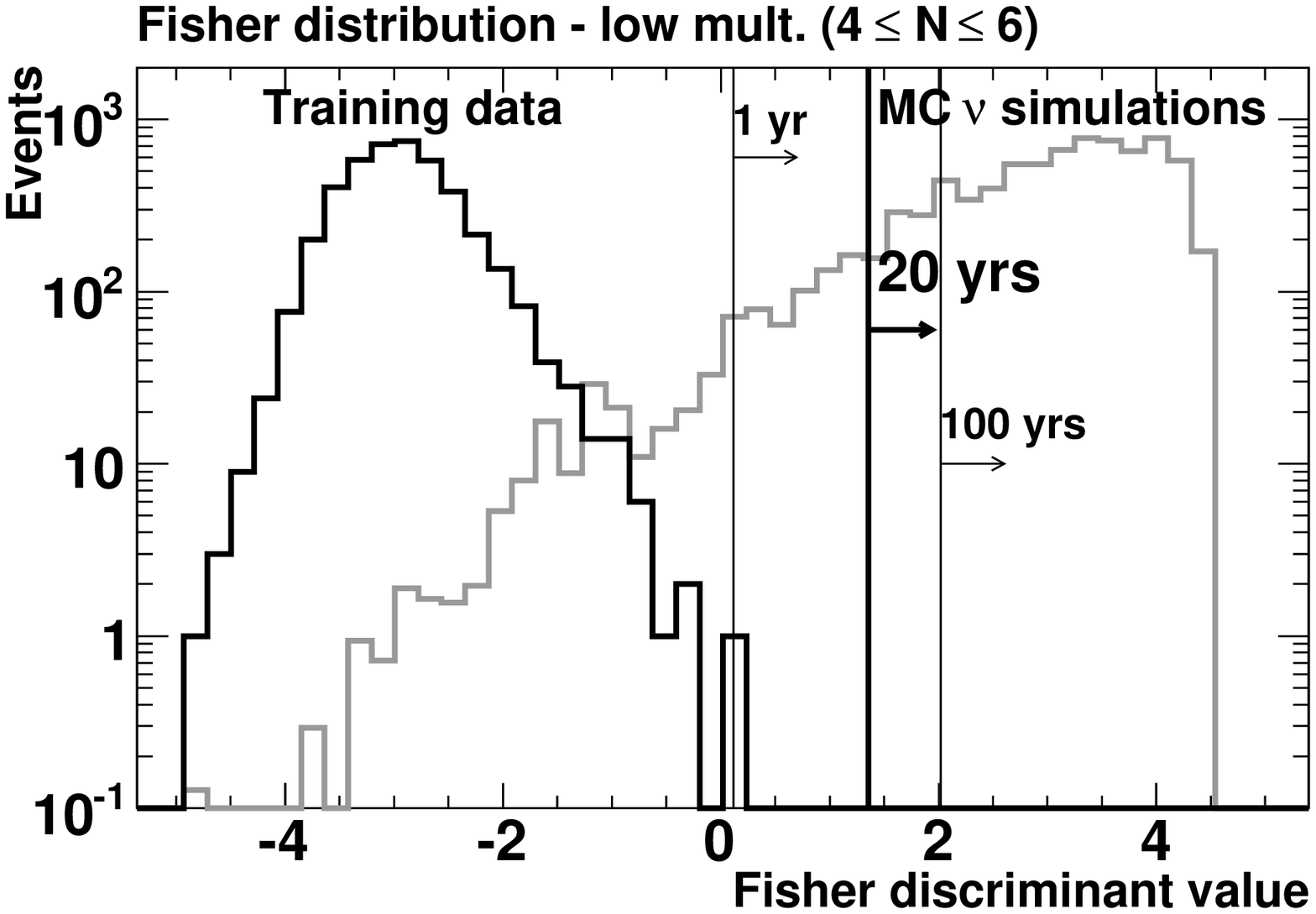} &
\includegraphics [width=8.6cm]{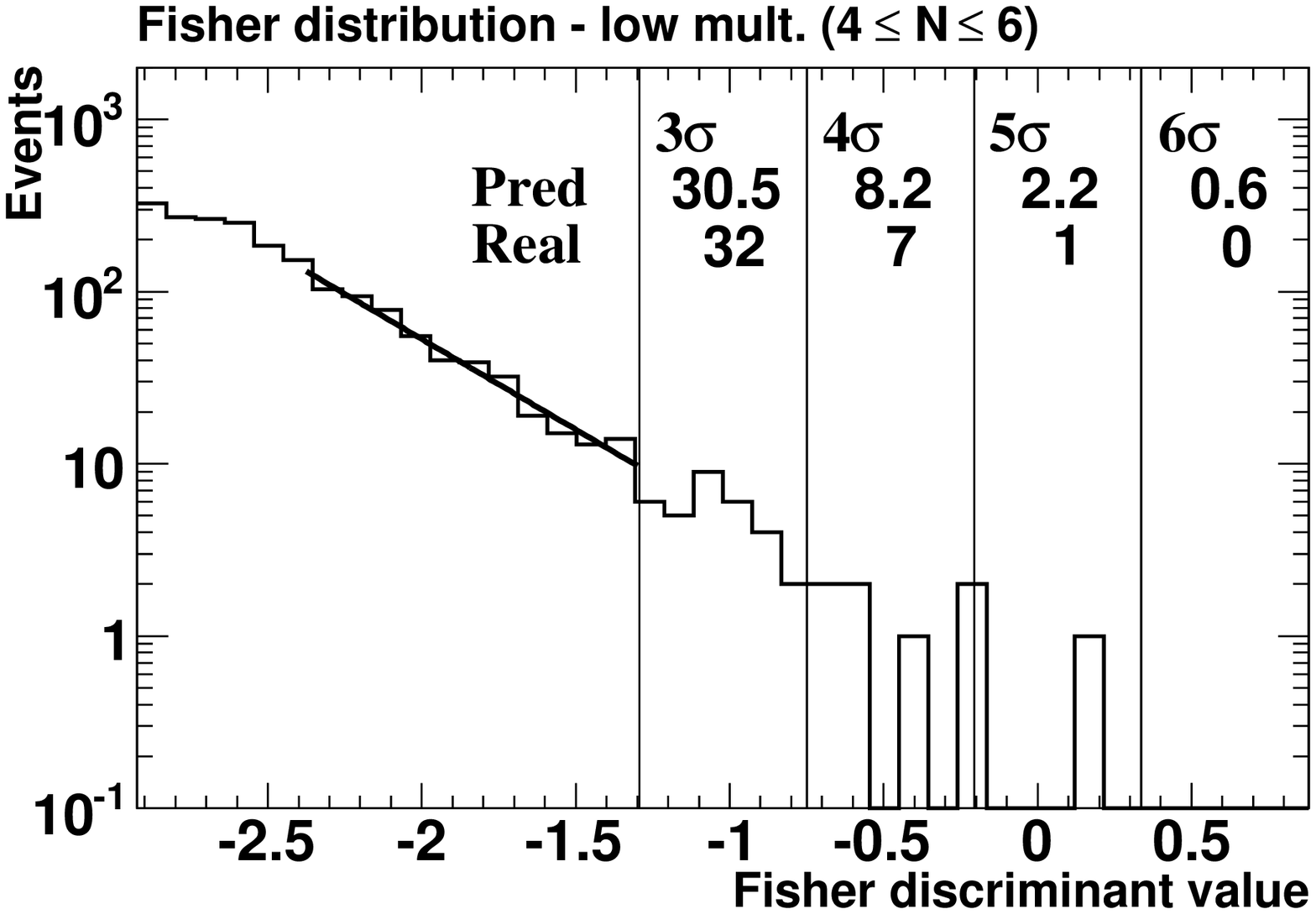} \\
\includegraphics [width=8.6cm]{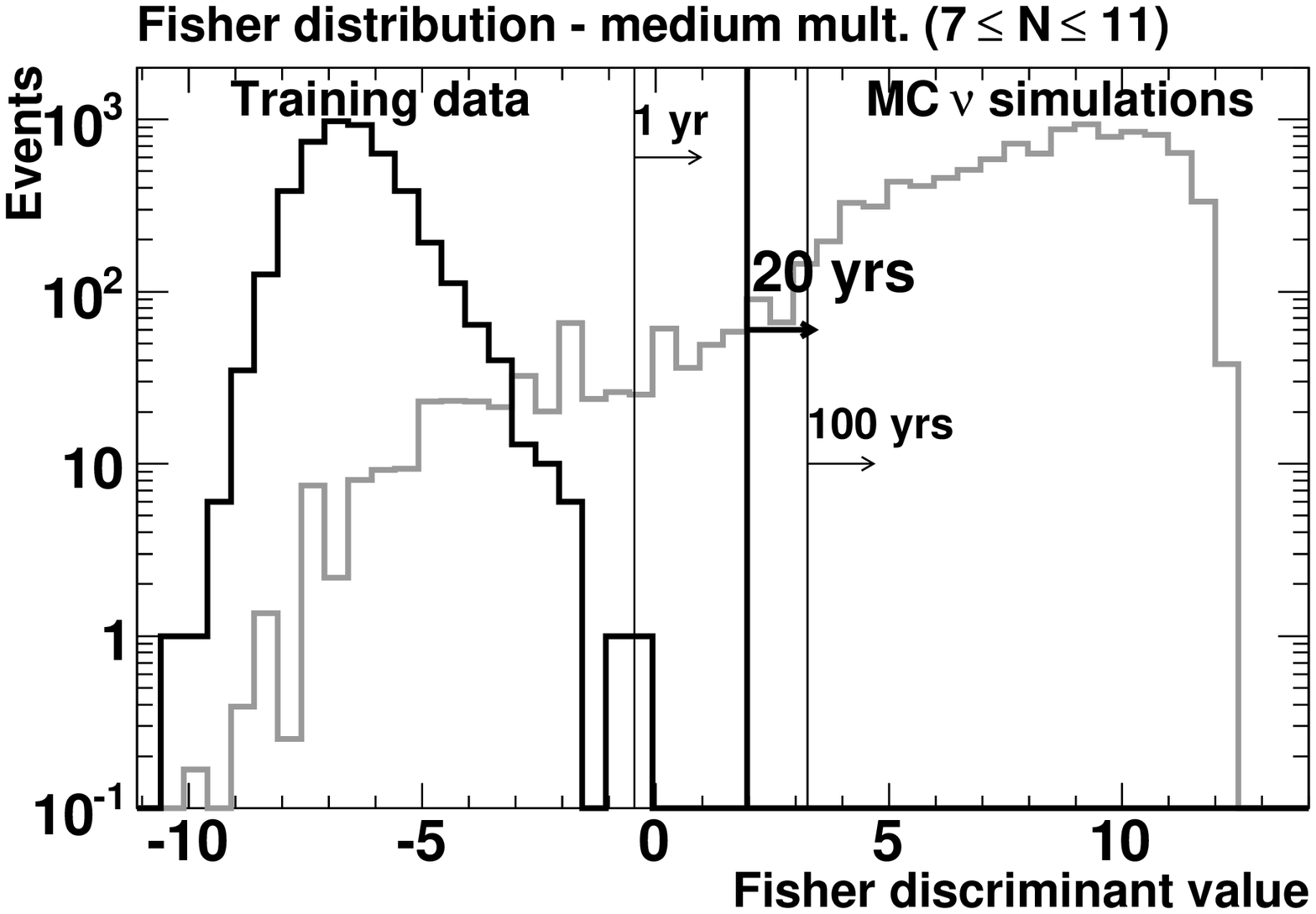} &
\includegraphics [width=8.6cm]{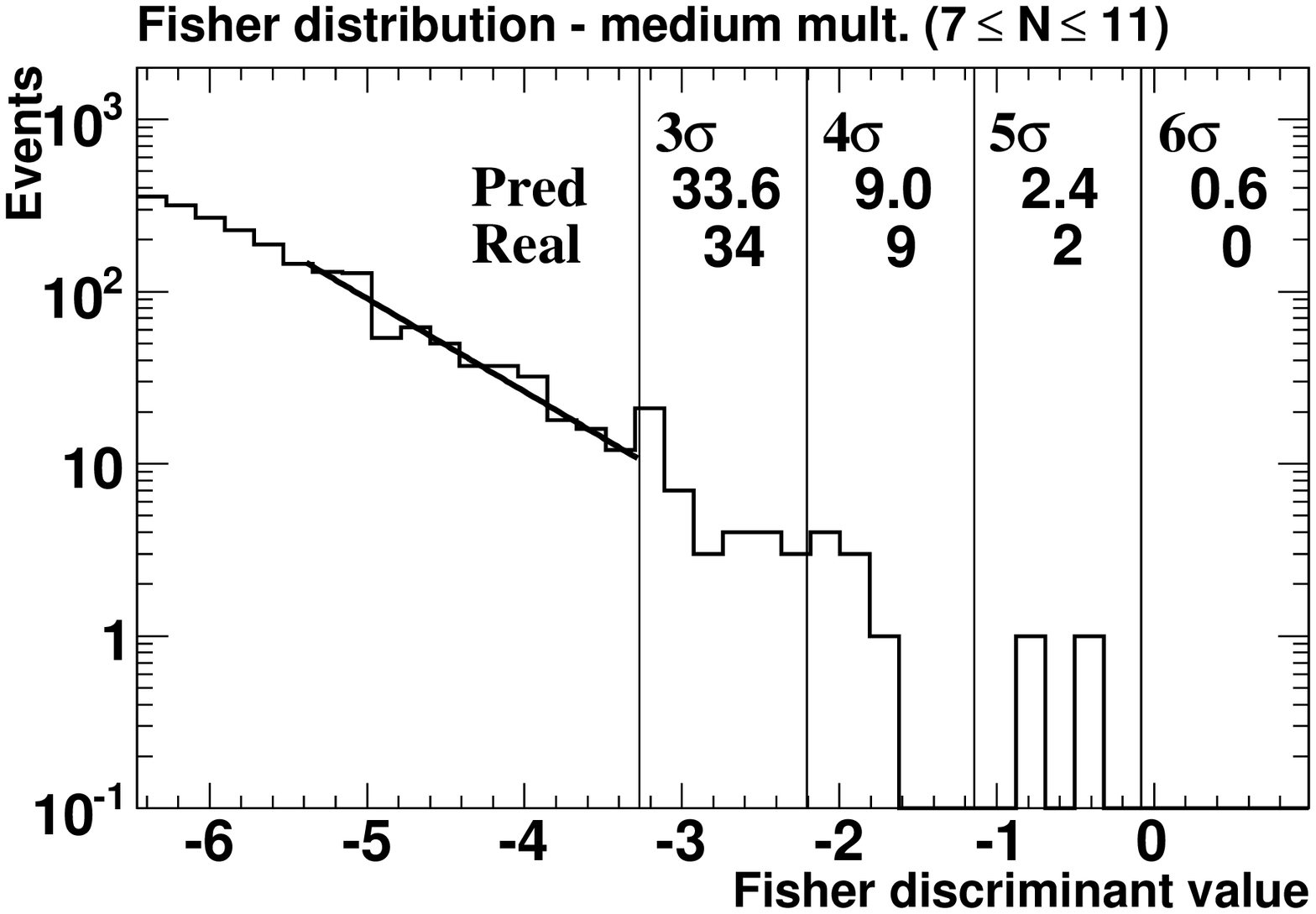} \\
\includegraphics [width=8.6cm]{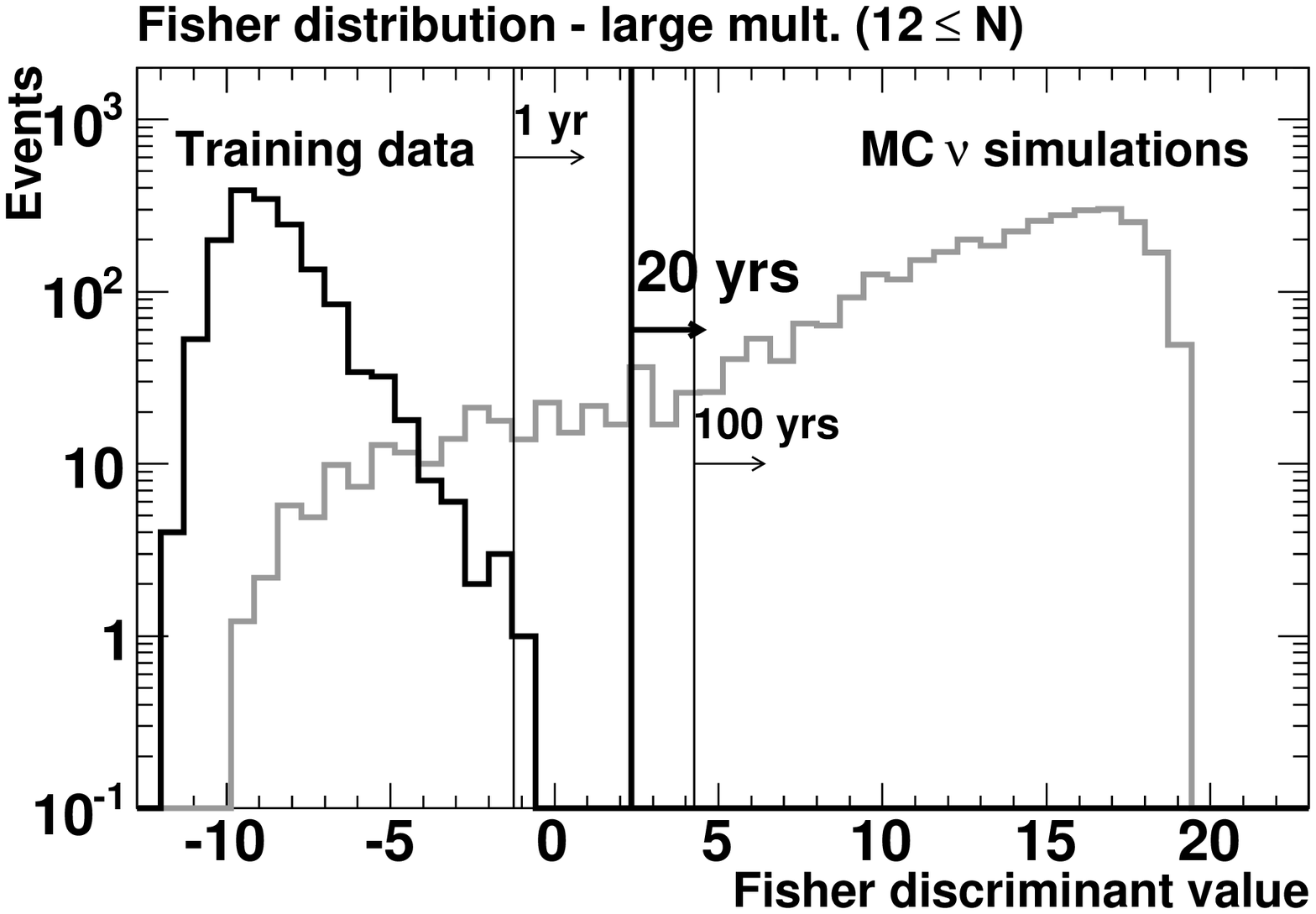} &
\includegraphics [width=8.6cm]{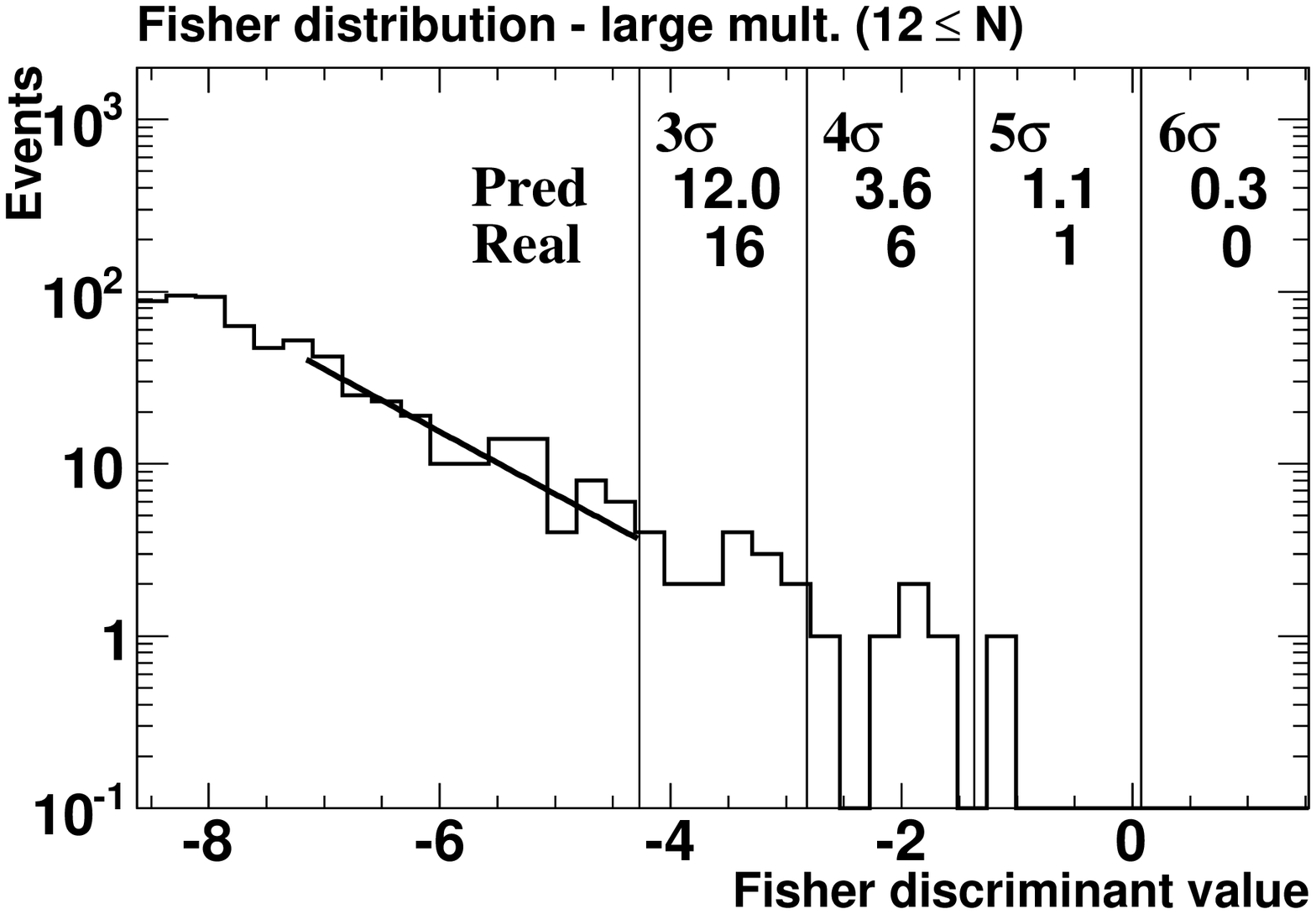} 
\end{array}$
\end{center}
\caption{\textit{Left panel:} distribution of the Fisher discriminant (see text for details) for events with station multiplicity $4 \leq N \leq 6$ (\textit{top}), 
$7 \leq N \leq 11$ (\textit{middle}), $12 \leq N$ (\textit{bottom}). Real data in the training period (1 Jan 04 - 31 Oct 07) describe the nucleonic background,
while Monte Carlo simulated down-going neutrinos correspond to the signal. The vertical lines indicate the cut in the Fisher value that needs 
to be placed to have less than 1 event in each period of time (1 yr, 20 yr, 100 yr). \textit{Right panel:} fit of an exponential function to the distribution 
of the Fisher discriminant ${\cal F}$ for the training data over the  $[1\sigma,3\sigma]$ interval. The predicted (Pred.), see text, and actual (Real) number 
of events are given for each of the test zones ($[3\sigma,4\sigma]$, $[4\sigma,5\sigma]$, $[5\sigma,6\sigma]$ and $[6\sigma,7\sigma]$).}
\label{fig:Fisher}
\end{figure*}

As the shower front is broader at larger distance from the core for both young
and old showers, the discrimination is better when splitting the samples
according to the multiplicity $N$ (number of selected stations). A Fisher
discriminant was built separately for $4\leq N \leq 6$, $7 \leq N \leq 11$ and
$N \geq 12$. Left panel of Fig.~\ref{fig:Fisher} shows the excellent separation achieved
for events in each of the 3 sub-samples.

Once the Fisher discriminant $\cal F$ is defined, one has to choose a threshold value that separates neutrino candidates from regular hadronic showers.
Because the predictions of the neutrino detection rates are very low, we want to keep the expected rate of background events incorrectly
classified as neutrinos well below any detectable signal: in practice, we wish it to be
less than one event for each multiplicity subsample within the expected 20-year lifetime of the Auger Observatory. 

The training period was used to produce a reasonable prediction of the background.
We observe that the tail of the background distribution of ${\cal F}$ is consistent with an exponential shape. In this way, we produced a fit to the distribution of ${\cal F}$ for the training data in the $[1\sigma,3\sigma]$ region, where $\sigma$ is the RMS of the training sample. This procedure is illustrated in Fig.~\ref{fig:Fisher}. We then
extrapolated it to find the cuts corresponding to 1 event per 1, 20 or 100 years on the full array.
The validity of the extrapolation is not guaranteed, but some physical arguments support an 
exponential tail, such as the fact that showers produced by nuclei or protons
(or even photons) have a distribution of $X_{max}$ that shows an
exponential shape, dictated by the distribution of the primary interaction.  
The exponential model may be checked below the cut by comparing the actual
number of events observed in the $[3\sigma,4\sigma]$, $[4\sigma,5\sigma]$, $[5\sigma,6\sigma]$ and $[6\sigma,7\sigma]$
regions, to the number of events predicted by extrapolating the fit done in the $[1\sigma,3\sigma]$ region. The values are in good agreement as shown in Fig.~\ref{fig:Fisher}.
For our search sample (equivalent to two years of full detector data) 
we have an estimated background of 0.1 events for each multiplicity class that add up to a total of 0.3 events with a statistical uncertainty of 30\%.
As we do not have at present a robust estimation of the background systematics we take a conservative approach and do not use this value to improve our flux upper limit.

As can be seen in Fig.~\ref{fig:Fisher} the identification cuts reject a small fraction of the neutrino events.
Consequently, its choice has only a small impact on the neutrino identification efficiency. 
The neutrino-induced showers rejected by these cuts are
those interacting far from the ground and similar to nucleonic-induced
showers.

\section{Identification efficiencies and Exposure}

During the data taking, the array was growing and had sporadic local inefficiencies.
Simulations of deep inclined neutrino showers indicate that besides an elongated pattern on the ground they have a large
longitudinal uncertainty on the core position. 
For these reasons we cannot apply (as done in the case of vertical showers~\cite{Auger_trigger}) a geometrical method relying on 
the estimated position of the shower core within a triangle or hexagon of active stations at each time.
Moreover, a shower can trigger the surface detector
even if the core falls outside the array. Besides, for deep inclined showers the
trigger and identification efficiencies depend not only on the shower energy
and zenith angle but also on the depth of the first interaction. For these
reasons a specific procedure was designed to compute the time dependent
acceptance and the integrated exposure.

The instantaneous status of the array is obtained from the trigger counting files,
which respond to the modifications of the array configuration at every second. 
To avoid having to cope with an enormous number of configurations,
we approximate the calculation of the aperture by subdividing 
the search period in three-day intervals, and we select a
reference array configuration to represent each. Once this is done
we calculate the neutrino identification efficiencies and the aperture assuming 
that the array remains unchanged during each three-day interval. 
Each reference configuration was chosen so that this approximation, 
if wrong, underestimates the exposure by a small amount ($\sim1$\%).

MC generated neutrino showers produced by {\sc aires} were randomly distributed over an extended circular area around the array, such that a shower
with a core falling outside this area has no chance to trigger the array. For
each three-day configuration the FADC traces of the active Cherenkov stations were
simulated, the local and global trigger conditions were applied and the events
were processed through the same reconstruction and identification algorithms
as the data (Sec.~\ref{sec:data_and_disc}).

\begin{figure}
\begin{center}
\includegraphics[width=8.6cm]{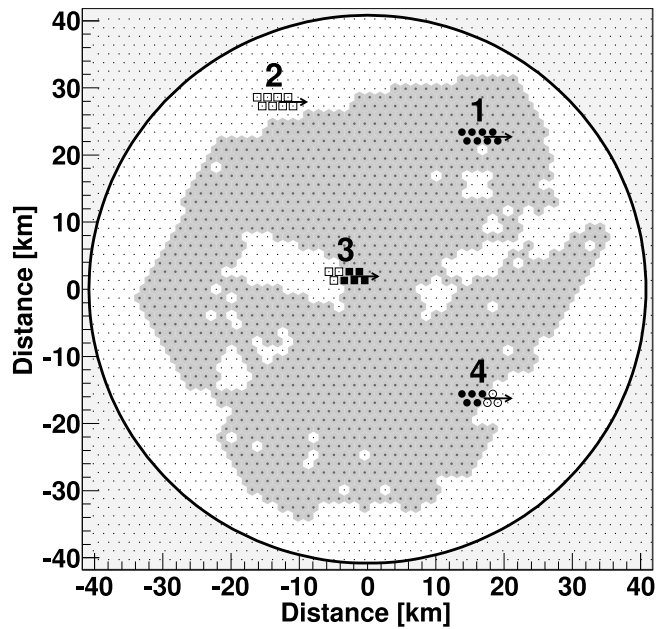}
\caption{An example of the result of placing the same deeply penetrating neutrino-induced shower 
at 4 different positions in an actual array configuration (shaded area) corresponding to 27 Oct 2007.
The arrows indicate the azimuthal arrival direction of the shower, the dots represent the infinite ideal array and the circumference the extended area (see text).
Solid symbols -- either circles or
squares -- correspond to triggered stations of the simulated shower that are also on the
actual array. Open symbols are stations that are not in the real array.
Shower 1 is completely contained and identified as a neutrino. Shower 2 falls entirely outside the real array and it does not trigger the array.
Although shower 3 triggers the array, it is not identified as a neutrino because the earliest three stations are not in the real array. 
Shower 4 loses some stations but keeps the earliest which are enough to identify the event as a neutrino.}
\label{Showers_in_Array}
\end{center}
\end{figure}

Fig.~\ref{Showers_in_Array} shows an example of a shower that would be a neutrino candidate
in an ideal array, placed at four random positions on the circular surface
defined above. Two of the realizations are effectively recognized as neutrino
events in the real array for that particular layout. The other two are either
not seen, or not identified as neutrinos.

\begin{figure}[ht]
\begin{center}
\includegraphics[width=8.6cm]{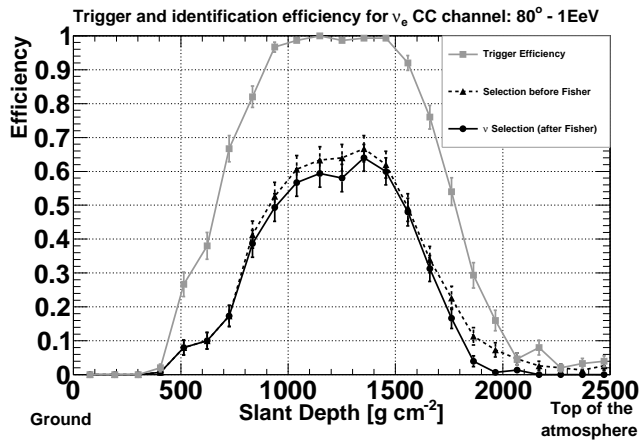}
\caption{Example of trigger and identification efficiency as a function of the slant depth of the interaction above the ground.
Notice that the Fisher discriminant neutrino selection actually keeps most of the neutrino showers that pass the quality and
reconstruction cuts discussed in Section~\ref{sec:eventSelection}.
}
\label{fig:effy_ideal}
\end{center}
\end{figure}

Fig.~\ref{fig:effy_ideal} shows the efficiency (fraction of events which pass
all steps) as a function of interaction depth in the atmosphere for neutrinos
of $E_\nu = 10^{18}$~eV and $\theta = 80^{\circ}$, in an ``ideal'' array without holes nor edges.
There is essentially a plateau between a minimal depth (needed for the
$\nu$-induced shower to reach a sufficient lateral expansion) and a maximal one
(such that the electromagnetic component is almost extinguished at ground
level) . Below and above this plateau, the efficiency drops rapidly to zero.
In other words, for a given channel and given values of $\theta$ and $E_{\nu}$, there is a
slice of atmosphere above the array where the interactions are detected and
distinguished: the matter contained in this volume will be referred to as the
``mass aperture'' in the following.


For each three-day period, we compute the effective area defined as the integral of the efficiency over core position:
\begin{equation}
A_{\rm eff}(E_\nu,\theta,D,t)=\int\!\varepsilon(\vec{r},E_\nu,\theta,D,t)\,{\rm d}A.
\end{equation}
The effective mass aperture $M_{\rm eff}(E_\nu,t)$ is obtained by integrating over the 
injection depth $D$ and the solid angle: 
\begin{equation}
M_{\rm eff}(E_\nu,t)=2\pi\int\!\!\!\int\,
\sin\theta\cos\theta \nonumber \,\,A_{\rm eff}(E_\nu,\theta,D,t)\,{\rm d}\theta\,{\rm d}D. \label{eq:Meff}
\end{equation}
To compute this integral we perform a spline interpolation on the finite three-dimensional 
mesh where $A_{\rm eff}$ is determined. The total mass aperture is then obtained summing 
$M_{\rm eff}(E_\nu,t)$ over different configurations corresponding to a certain period of time.
It is defined independently of the $\nu$-nucleon cross-section.

\begin{figure}
\begin{center}
\noindent
\includegraphics [width=8.6cm]{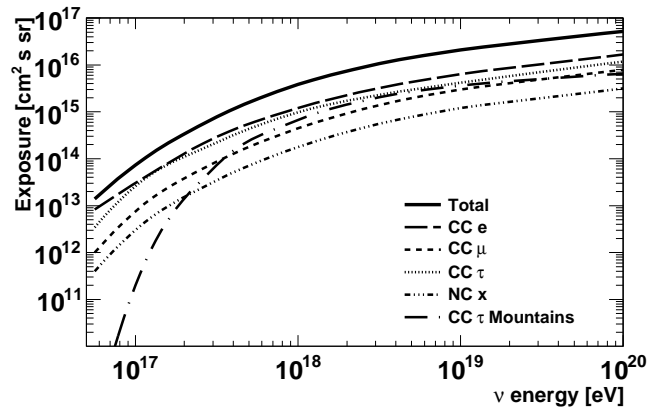}
\end{center}
\caption{SD Exposure for our search period for
down-going neutrino-initiated showers. The total exposure is shown as a full line. The exposure for
individual neutrino flavours and interaction channels is also shown.
}
\label{fig:Meff_vs_E}
\end{figure}

A combined exposure can be obtained by a summation over the search period:
\begin{equation}\label{eq:exposure}
{\cal E}(E_\nu)=\sum_i\left[{\omega^i\,\sigma^i(E_\nu)\int\!\frac{M^i_{\rm eff}(E_\nu,t)}{m}\,{\rm d}t}\right].
\end{equation}
The sum runs over the three neutrino flavours (with fractions $\omega^i$) and the
CC and NC interactions; $m$ is the mass of a nucleon. Here we
assume a full $\nu_{\tau} \leftrightarrow \nu_{\mu}$ mixing, leading
to $\omega^i=1$ for the three flavours.

We use the $\nu$--nucleon cross-section given in~\cite{cooper_sarkar} (CSS hereafter) to
compute the reference exposure of our search period. It is shown in Fig.~\ref{fig:Meff_vs_E} as a function of neutrino energy. 
In Table~\ref{tab:effMass} we also give the mass aperture integrated in time for all the considered channels,
allowing the reader to compute the exposure using different cross-sections or flux models.
\begingroup
\squeezetable
\begin{table}[ht]
\begin{center}
\begin{tabular}{cccccc}
\hline \hline \noalign{\smallskip}
\boldmath{$\log E/\mbox{\textbf{eV}}$} & \boldmath{$\nu_e$} \textbf{CC} & \boldmath{$\nu_{\mu}$} \textbf{CC} & \boldmath{$\nu_{\tau}$} \textbf{CC} & \boldmath{$\nu_x$} \textbf{NC}  & \boldmath{$\nu_{\tau}$} \textbf{Mount.} \\
\hline \noalign{\smallskip}
$16.75$ & $4.35\cdot 10^{21}$ & $5.27\cdot10^{20}$ & $1.82\cdot10^{21}$ & $2.11\cdot10^{20}$ & -\\
$17$ & $1.27\cdot10^{22}$ & $3.16\cdot10^{21}$ & $1.09\cdot10^{22}$ & $1.26\cdot10^{21}$ & -\\
$17.5$ & $7.94\cdot10^{22}$ & $2.34\cdot10^{22}$ & $6.02\cdot10^{22}$ & $9.37\cdot10^{21}$ & $1.98\cdot10^{22}$\\
$18$ & $2.17\cdot10^{23}$ & $8.01\cdot10^{22}$ & $1.77\cdot10^{23}$ & $3.20\cdot10^{22}$ & $1.21\cdot10^{23}$\\
$18.5$ & $3.95\cdot10^{23}$ & $1.71\cdot10^{23}$ & $2.84\cdot10^{23}$ & $6.84\cdot10^{22}$ & $2.51\cdot10^{23}$\\
$19$ & $5.44\cdot10^{23}$ & $2.56\cdot10^{23}$ & $3.58\cdot10^{23}$ & $1.03\cdot10^{23}$ & $3.13\cdot10^{23}$\\
$19.5$ & $6.32\cdot10^{23}$ & $2.99\cdot10^{23}$ & $4.36\cdot10^{23}$ & $1.20\cdot10^{23}$ & $3.06\cdot10^{23}$\\
$20$ & $7.29\cdot10^{23}$ & $3.45\cdot10^{23}$ & $5.19\cdot10^{23}$ & $1.38\cdot10^{23}$ & $2.82\cdot10^{23}$\\
\hline \hline
\end{tabular}
\caption{\label{tab:effMass}
Effective mass aperture integrated over time for the search period (1 Nov 2007 to 31 May 2010) for down-going neutrinos
of the Pierre Auger Surface Detector (in units of [g~sr~s]).}
\end{center}
\end{table}
\endgroup

\section{Systematic uncertainties}
\label{sec:syst}

The calculation of the mass aperture of the Auger Observatory for neutrino showers requires the
input of several ingredients which we have selected from amongst conventionally used
options.  Some of these choices are directly related to the
Monte Carlo simulation of the showers, i.e. generator of the first neutrino interaction, 
parton distribution function (PDF), air shower development and hadronic
model. Others have to do with the precision of our knowledge of the topography
of the mountains surrounding the Observatory, and some come from the
limitations on the theoretical models estimating, for instance, the interaction
cross-section or the $\tau$ energy loss at high energies.
By adding linearly all these contributions, our estimate of the total systematic uncertainty on the exposure amounts to
+22\% -46\%.

In the following subsections we discuss in detail the dependence of the exposure on each of the above mentioned choices by modifying 
the different ingredients one by one.

\subsection{Monte Carlo simulation of the shower}

The reference Monte Carlo neutrino sample was produced with {\sc herwig} 6.5.10~\cite{HERWIG} as interaction
generator in combination with the CTEQ06m~\cite{cteq} parton distribution functions,
AIRES 2.8 as shower simulator (thinning value of $10^{-6}$) and QGSJETII.03~\cite{QGSJETII}
as hadronic model.

\begin{figure}[floatfix]
\centering
\includegraphics[width=8.6cm]{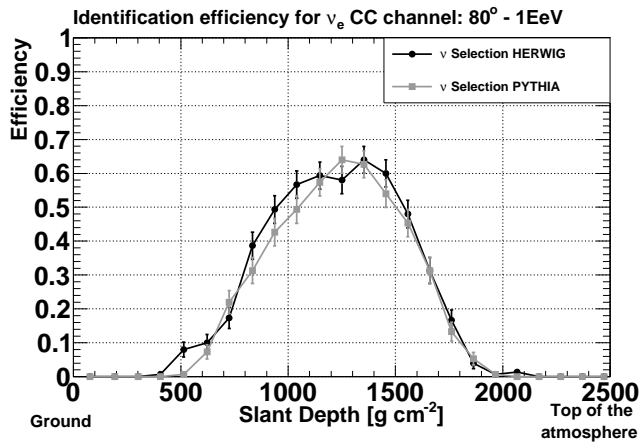}
\caption{Identification efficiency
as function of the slant depth for systematic uncertainties studies.
Comparison between interaction generators ({\sc herwig} and {\sc pythia}). The rest of the Monte Carlo input parameters remain the same.
}
\label{fig:qgs_sib}
\end{figure}

In order to assess the influence of this particular choice of models
on the detector aperture, independent sets of CC $\nu_{e}$ showers were generated
at 1~EeV and 80$^\circ$ using
different combinations of several interaction generators, PDFs,
shower simulators, thinning values and hadronic models.
We chose this particular energy and angle bin because it is 
the one that contributes the most to the limit.

In Fig.~\ref{fig:qgs_sib} we show, as an example, the detection efficiency
as a function of the slant depth when using our reference options ({\sc herwig}) and when changing only the interaction generator ({\sc pythia}).
Since the shapes of the neutrino-selection efficiency curves remain similar, we can estimate the effect of changing the interaction generator by computing the 
integral of the curves and reporting the relative difference (RD) between them. The same procedure is applied to estimate the effect of changing other ingredients of the simulation. 
A summary of this RDs is given in Table~\ref{tab:mc_sys}.

\begingroup
\squeezetable
\begin{table}[floatfix]
\begin{center}
    \begin{tabular}{lll@{}c@{}}
\hline\hline \noalign{\smallskip}
\textbf{Parameter}      & \textbf{Reference}   & \textbf{Modification} & \textbf{RD}\\ \noalign{\smallskip}
                        & \textbf{(A)}         & \textbf{(B)} & $\mathbf{\frac{\int B\,-\,\int A}{(\int B + \int A)/2}}$\\ \noalign{\smallskip}
\hline\noalign{\smallskip}
 Interaction generator  & HERWIG  & PYTHIA~\cite{PYTHIA}     & -7\%\\
                        &         & HERWIG++~\cite{Herwig++}   & -7\%\\
\hline\noalign{\smallskip}
 PDF (gen. level)       & CTEQ06m & MSTW~\cite{mstw}      & -7\%\\                                  
\hline\noalign{\smallskip}
 Shower Simulator       & AIRES   & CORSIKA 6.9~\cite{corsika}    & -17\%\\
\hline\noalign{\smallskip}
 Hadronic Model         & QGSJETII & QGSJETI~\cite{QGSJET}  & +2\%\\
                        &          & SIBYLL~\cite{SIBYLL}   & -2\%\\
                        &          & SIBYLL ($E$=0.3~EeV)   & -1\%\\
                        &          & SIBYLL ($E$=3~EeV)     & -2\%\\
                        &          & SIBYLL ($\theta$=85$^\circ$)  & 0\%\\
                        &          & SIBYLL ($\theta$=89$^\circ$)  & +4\%\\
\hline
\hline\noalign{\smallskip}
 Thinning                & $10^{-6}$ &$10^{-7}$& +7\%\\
\hline\hline 
\end{tabular}  
  \end{center}
\caption{
\label{tab:mc_sys}
Summary of the relative differences (RD) between the reference calculation of the exposure and the calculations done changing one of the ingredients of the Monte Carlo simulations at a time. The RD were obtained for zenith angle $\theta$=80$^\circ$ and energy $E$=1~EeV unless otherwise stated.
The statistical uncertainty of all the relative differences is $\pm$4\%.
}
\end{table}
\endgroup

We observe that the changes in interaction generator, PDF, shower simulator and hadronic model brought about a decrease of the estimated aperture, with the choice of the shower simulation being the dominating effect. 
On the other hand, an improvement of the relative thinning level causes the opposite effect. 
Although we cannot recompute the aperture for all possible alternatives of the relevant ingredients, the relative differences reported in Table~\ref{tab:mc_sys} serve as an estimate of the systematic dependence of our result on the simulation options.
For each category of potential systematic effects in Table~\ref{tab:mc_sys}, we take the
maximum observed RD as an estimate of the corresponding systematic uncertainty.
A total systematic uncertainty of +9\% -33\% on the exposure is obtained by linear
addition of the maximum positive and negative deviations.

\subsection{$\nu$--nucleon cross-sections and $\tau$ energy loss}

We adopted the uncertainty in the $\nu$-nucleon cross-section as calculated in~\cite{cooper_sarkar}. 
It translates into a $\pm$7~\% uncertainty in the total exposure.
In any case, as mentioned above, Table~\ref{tab:effMass} shows the Auger mass aperture for 
down-going neutrinos which does not depend on the $\nu$ cross-section; hence the 
expected neutrino event rate (and neutrino flux limit) can be computed as necessary for 
other models and values of the $\nu$ cross-section (see e.g.~\cite{Connolly, newSarkar}).

\subsection{Topography}
 As explained in section~\ref{sec:mountains}, the actual topography surrounding
the Observatory has been taken into account by detailed Monte Carlo simulations
which include digital elevation maps. In principle, uncertainties due to different
tau energy loss models should not be important for down-going neutrinos, but, due to
the fact that the Pierre Auger Observatory is close to the Andes, a non negligible
contribution to the event rate from down-going $\tau$ neutrinos interacting in the
mountains and producing a $\tau$ lepton is expected (see Table~\ref{tab:limit_disc}).
The systematic error on the total reference exposure due to this channel amounts to $\pm$6\%, 
dominated by the uncertainties on the cross-section and energy loss models.

\section{Results and discussion}
\label{sec:results}
In this section we present the calculation of the upper limit to the diffuse flux of UHE$\nu$s
and compare our results to some selected model predictions and discuss the implications.

\subsection{Upper limit on the diffuse neutrino flux}
\label{sec:limit}
Once the multivariate algorithms and selection cuts defining a neutrino candidate were studied and tuned with the
Monte Carlo simulations and the training data sample, we applied them to the search data sample.
We first tested the compatibility between the shapes of the tails of the Fisher distributions during training and search periods 
by using an unbinned Kolmogorov hypothesis test, and found them to be in agreement with p-values of 0.37, 0.16 and 0.17 for the small, medium and large multiplicity classes, respectively.
\begin{figure}[floatfix]
\includegraphics[width=8.6cm]{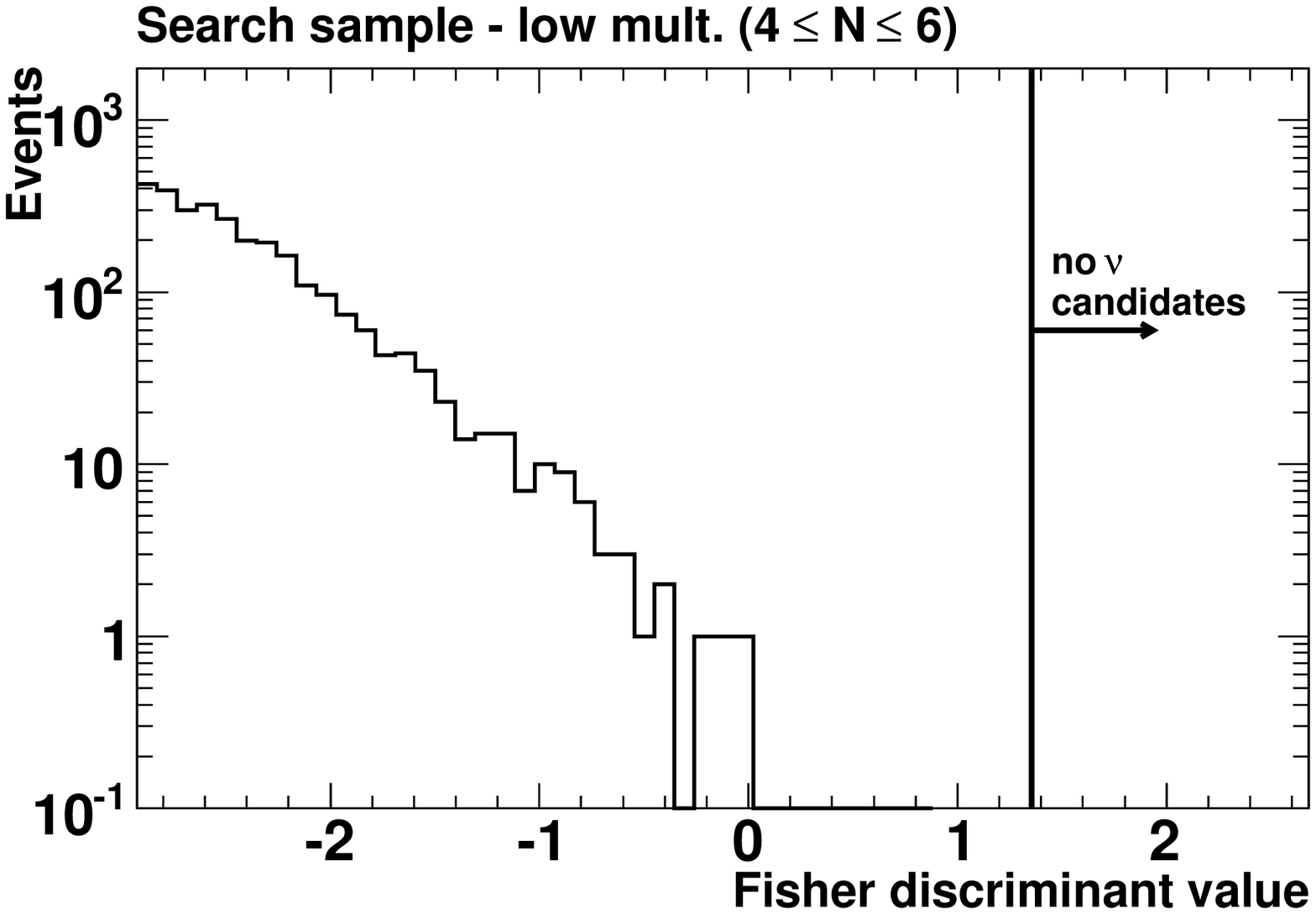} \\
\includegraphics[width=8.6cm]{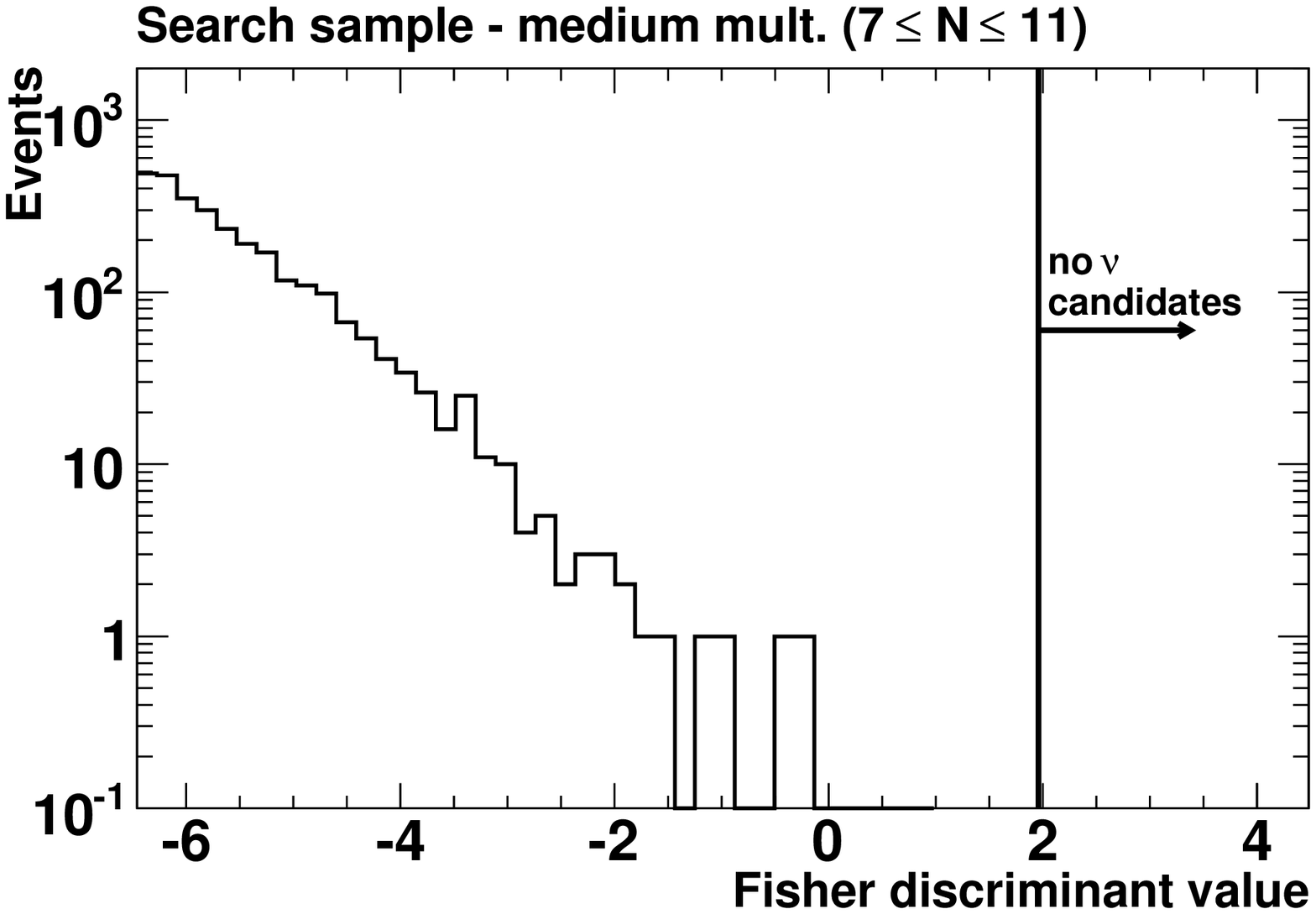} \\
\includegraphics[width=8.6cm]{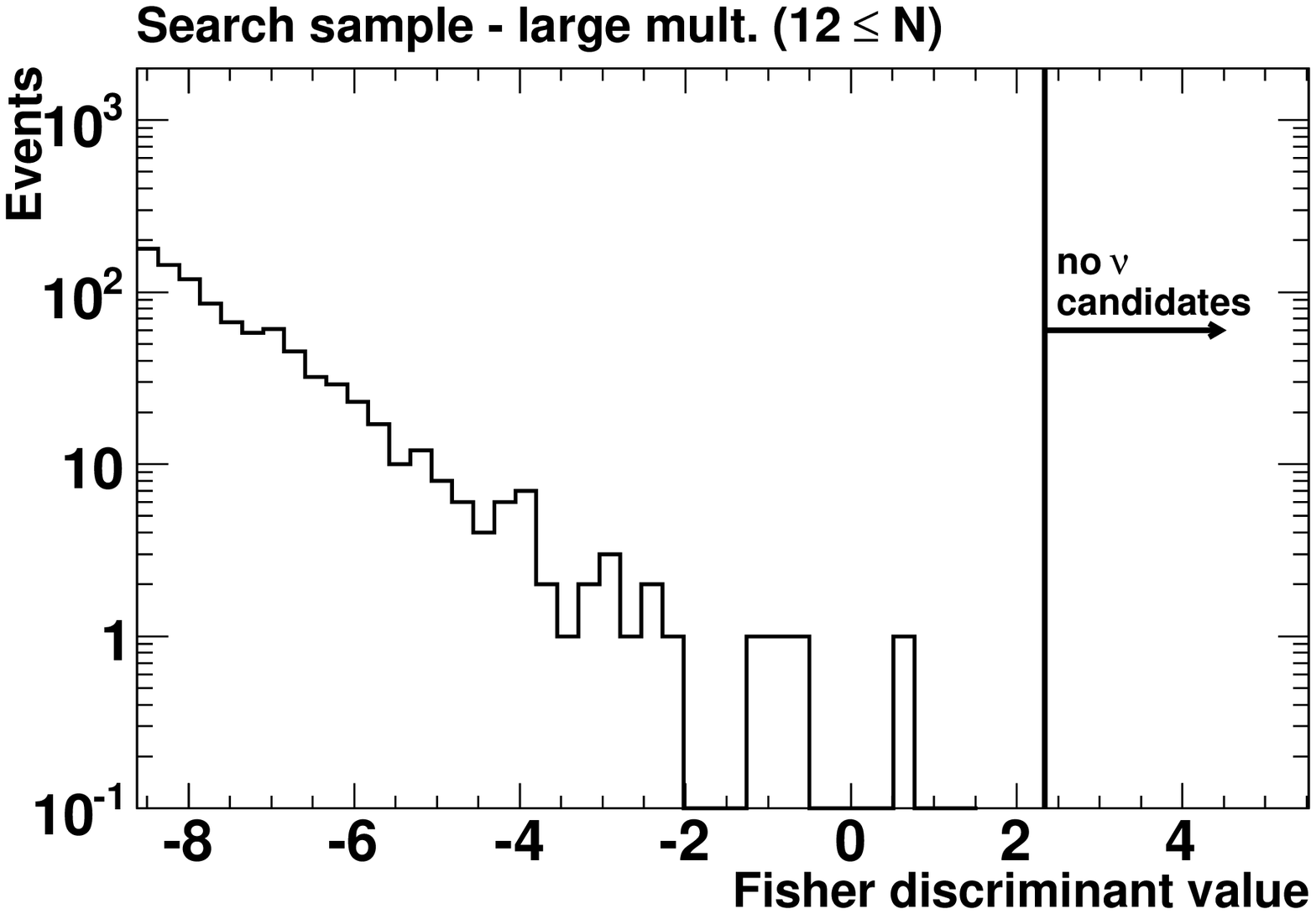} 
\caption{Fisher distribution of the search sample (1 Nov 07 - 31 May 10, 2010) for events with multiplicity $4 \leq N \leq 6$ (\textit{top}), 
$7 \leq N \leq 11$ (\textit{middle}), $12 \leq N$ (\textit{bottom}). No neutrino candidates are found.}
\label{fig:searchTail}
\end{figure}

We found no candidate events in the search period (see Fig.~\ref{fig:searchTail}). The highest test zone with events in the Fisher distribution of the search sample is the 6-7 sigma region.
It has only one event and we expected 2.2 from the exponential fit to the test sample.

\begin{figure}
\begin{center}
\noindent
\includegraphics [width=8.6cm]{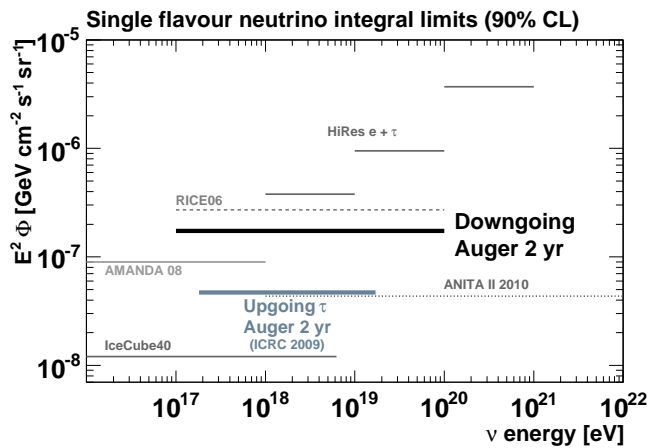}
\end{center}
\caption{
Integrated upper limits (90$\%$ CL)
from the Pierre Auger Observatory 
for a diffuse flux of down-going $\nu$ in the period 1 Nov 2007 - 31 May 2010. 
For comparison up-going $\nu_\tau$ (1 Jan 2004 - 28 Feb 09)\cite{Tiffenberg_icrc09} and limits from other
experiments~\cite{nu_limits} are also plotted.
}\label{fig:intLimits}
\end{figure}

The expected number of events from a diffuse flux of neutrinos in a given energy
range is given by:
\begin{equation}
N_{\rm expected}=\int_{E_{\rm min}}^{E_{\rm max}}\Phi(E_{\nu})\,{\cal E}(E_{\nu})\,{\rm d}E_{\nu},
\end{equation}
where ${\cal E}(E_{\nu})$ is our reference exposure (eq.~\ref{eq:exposure} and Fig.~\ref{fig:Meff_vs_E}).
The upper limit is derived for a differential neutrino flux $\Phi(E_\nu) = k\cdot E_\nu^{-2}$. Also, we
assume that due to neutrino oscillations the diffuse flux is composed of electron,
muon and $\tau$ neutrinos in the same proportion. We expect less than one background event
after the neutrino selection procedure is applied to the data sample corresponding
to the reference exposure (see section 5). Given the uncertainties of this estimate,
the number of background events will be assumed to be zero, which results in a
more conservative upper limit. A semi-Bayesian extension~\cite{Conrad_limit} of the Feldman-
Cousins approach~\cite{Feldman-Cousins} is used to include the uncertainty in the exposure, giving an
upper limit at 90\% CL on the integrated flux of diffuse neutrinos of:
\begin{equation}
k < 1.74 \times 10^{-7}~{\rm GeV~cm^{-2}~s^{-1}~sr^{-1}}.
\end{equation}
The effect of including the systematics from MC, $\nu$--nucleon cross-sections and $\tau$ energy loss is to increase the limit by 15\%.
The limit is quoted for a single neutrino flavour. The
relative importance of each neutrino flavour in the determination of the upper limit
can be derived from Table~\ref{tab:limit_disc}, which gives the expected fractions of neutrinos in
the selected sample according to their flavour and interaction channel. The largest
contribution comes from $\nu_e$~CC. The second largest is $\nu_{\tau}$~CC, due to double-bang
interactions and the large average fraction of energy going into the shower in the
decay of the $\tau$ lepton. Our result together with other experimental limits~\cite{nu_limits} is shown in Fig.~\ref{fig:intLimits}.
\begin{table}[ht]
\begin{center}
\begin{tabular}{lccc }
\hline \hline
 \textbf{Channel}  & \textbf{CC}   & \textbf{NC} & \textbf{Total}\\
\hline \noalign{\smallskip}
 {$e$             }        & 33\%   & 5\%  & 38\% \\
 {$\mu$           }        & 13\%   & 5\%  & 18\% \\
 {$\tau$ air      }        & 24\%   & 5\%  & 29\% \\
 {$\tau$ mountains}        & 15\%   &      & 15\% \\
\hline \noalign{\smallskip}
 {\bf Total}               & 85\%   & 15\% & 100\% \\
\hline \hline
\end{tabular}
\caption{\label{tab:limit_disc}
Expected fractions of neutrinos in the selected sample according to their flavour and interaction channel
(CC and NC). These fractions are derived assuming that electron, muon and $\tau$
neutrinos are in the same proportion in the diffuse flux.
}
\end{center}
\end{table}

\begin{figure}
\begin{center}
\noindent
\includegraphics [width=8.6cm]{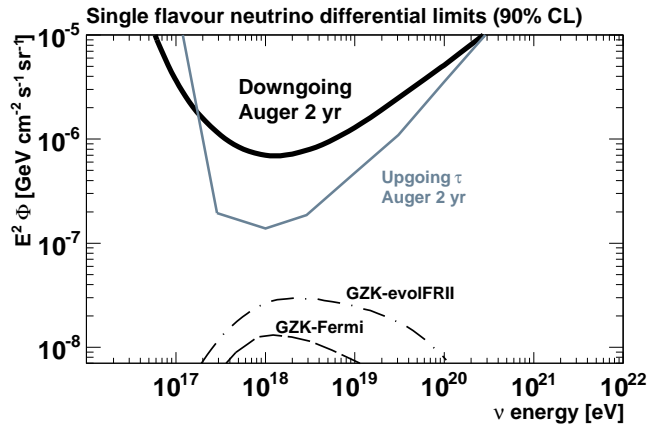}
\end{center}
\caption{
Differential limits (90$\%$ CL)
from the Pierre Auger Observatory 
for a diffuse flux of down-going $\nu$ in the period 1 Nov 2007 - 31 May 2010 
and up-going $\nu_\tau$ (1 Jan 2004 - 28 Feb 2009)\cite{Tiffenberg_icrc09}.
For reference, two recent calculations of this flux are shown: ``GZK-Fermi''~\cite{nu_GZK_Ahlers} takes into
account the Fermi-LAT constraint on the GZK cascade photons; the other 
``GZK-evolFRII''~\cite{nu_GZK_Kotera} adopts a strong source evolution model for FR-II galaxies, assumed to be the sources of UHECRs.
}\label{fig:difLimits}
\end{figure}

Another usual way of presenting the upper bound is in the less-model-dependent differential form. 
It assumes that the diffuse neutrino flux behaves as $\sim1/E^2$ within energy bins of unity width 
on a natural logarithmic scale, and is given by $2.44 \slash{\cal E}(E_{\nu})E_{\nu}$ 
accounting for statistical uncertainties only and assuming no background~\cite{AnchoDiffLimit}.
The differential limit obtained including systematic uncertainties is shown in Fig.~\ref{fig:difLimits}, together with our previous result on up-going $\nu_\tau$~\cite{Tiffenberg_icrc09} and two theoretical predictions for cosmogenic neutrinos~\cite{nu_GZK_Ahlers, nu_GZK_Kotera}. We observe that we achieve maximum sensitivity in the 0.3-10~EeV energy range.

\subsection{Model predictions}
\label{sec:models}

There is a wide variety of models predicting fluxes of neutrinos with energies in the EeV range~\cite{nu_reviews}. They are usually separated into three groups: cosmogenic neutrinos~\cite[e.g.][]{nu_GZK_Ahlers,nu_GZK_Kotera}, neutrinos produced in accelerating sources~\cite[e.g.][]{MPR01, BBR} and neutrinos of exotic origin~\cite[e.g.][]{Sigl}. In all these models there are parameters with unknown values which can change the spectral shape and strength of the flux. 
In Table~\ref{tab:theoreticalModels} we give the event rates after folding these fluxes with our reference exposure.
\begin{table}
\begin{center}
\begin{tabular}{lcc}
\hline \hline \noalign{\smallskip}
\textbf{Reference} & \boldmath{$N$} \textbf{expected} & \textbf{P. of obs. 0}\\
\hline \noalign{\smallskip}
GZK-Fermi & 0.1 &  0.9 \\ 
GZK-evolFRII & 0.3 &  0.7\\ 
\hline\noalign{\smallskip}
MPR-max  & 2.0 &   0.1\\ 
BBR  & 0.8 &    0.4\\ 
\hline\noalign{\smallskip}
TD-Necklaces & 0.8 &  0.4\\ 
Z-Burst  & 7.8 &   $4\times10^{-4}$\\ 
\hline \hline
\end{tabular}
\vspace{5mm}
\caption{Expected number of events using the current exposure of
down-going $\nu$ measured by the Pierre Auger Observatory for several models~\cite{nu_GZK_Ahlers,nu_GZK_Kotera, MPR01, BBR, Sigl}. The third column gives the probabilities of observing 0 events given that we expect $N$.} 
\label{tab:theoreticalModels}
\end{center}
\end{table}

Current theoretical flux predictions for cosmogenic neutrinos~\cite{nu_GZK_Ahlers,nu_GZK_Kotera} seem to be out of reach of our present sensitivity. Concerning neutrinos produced in accelerating sources there are popular models~\cite{MPR01, BBR} which predict event rates which could be detected in the next few years. Regarding exotic models~\cite{Sigl}, TD-Necklaces will be within our sensitivity range in one or two  years, while Z-Burst models are
already strongly disfavoured. Note that all such `top down' models are also tightly constrained by the limits of the Pierre Auger Observatory on the photon fraction in
UHECR~\cite{Auger_photon_limit}.

\section{Acknowledgments}
%
The successful installation, commissioning and operation of the Pierre Auger Observatory
would not have been possible without the strong commitment and effort
from the technical and administrative staff in Malarg\"ue.

We are very grateful to the following agencies and organizations for financial support: 
Comisi\'on Nacional de Energ\'ia At\'omica, 
Fundaci\'on Antorchas,
Gobierno De La Provincia de Mendoza, 
Municipalidad de Malarg\"ue,
NDM Holdings and Valle Las Le\~nas, in gratitude for their continuing
cooperation over land access, Argentina; 
the Australian Research Council;
Conselho Nacional de Desenvolvimento Cient\'ifico e Tecnol\'ogico (CNPq),
Financiadora de Estudos e Projetos (FINEP),
Funda\c{c}\~ao de Amparo \`a Pesquisa do Estado de Rio de Janeiro (FAPERJ),
Funda\c{c}\~ao de Amparo \`a Pesquisa do Estado de S\~ao Paulo (FAPESP),
Minist\'erio de Ci\^{e}ncia e Tecnologia (MCT), Brazil;
AVCR AV0Z10100502 and AV0Z10100522,
GAAV KJB100100904,
MSMT-CR LA08016, LC527, 1M06002, and MSM0021620859, Czech Republic;
Centre de Calcul IN2P3/CNRS, 
Centre National de la Recherche Scientifique (CNRS),
Conseil R\'egional Ile-de-France,
D\'epartement  Physique Nucl\'eaire et Corpusculaire (PNC-IN2P3/CNRS),
D\'epartement Sciences de l'Univers (SDU-INSU/CNRS), France;
Bundesministerium f\"ur Bildung und Forschung (BMBF),
Deutsche Forschungsgemeinschaft (DFG),
Finanzministerium Baden-W\"urttemberg,
Helmholtz-Gemeinschaft Deutscher Forschung\-szen\-tren (HGF),
Ministerium f\"ur Wissenschaft und Forschung, Nordrhein-Westfalen,
Ministerium f\"ur Wissenschaft, Forschung und Kunst, Baden-W\"ur\-ttemberg, Germany; 
Istituto Nazionale di Fisica Nucleare (INFN),
Ministero dell'Istruzione, dell'Universit\`a e della Ricerca (MIUR), Italy;
Consejo Nacional de Ciencia y Tecnolog\'ia (CONACYT), Mexico;
Ministerie van Onderwijs, Cultuur en Wetenschap,
Nederlandse Organisatie voor Wetenschappelijk Onderzoek (NWO),
Stichting voor Fundamenteel Onderzoek der Materie (FOM), Netherlands;
Ministry of Science and Higher Education,
Grant Nos. 1 P03 D 014 30, N202 090 31/0623, and PAP/218/2006, Poland;
Funda\c{c}\~ao para a Ci\^{e}ncia e a Tecnologia, Portugal;
Ministry for Higher Education, Science, and Technology,
Slovenian Research Agency, Slovenia;
Comunidad de Madrid, 
Consejer\'ia de Educaci\'on de la Comunidad de Castilla La Mancha, 
FEDER funds, 
Ministerio de Ciencia e Innovaci\'on and Consolider-Ingenio 2010 (CPAN),
Xunta de Galicia, Spain;
Science and Technology Facilities Council, United Kingdom;
Department of Energy, Contract Nos. DE-AC02-07CH11359, DE-FR02-04ER41300,
National Science Foundation, Grant No. 0450696,
The Grainger Foundation USA; 
ALFA-EC / HELEN,
European Union 6th Framework Program,
Grant No. MEIF-CT-2005-025057, 
European Union 7th Framework Program, Grant No. PIEF-GA-2008-220240,
and UNESCO.

\end{document}

%% file: auger_authors_RevTex_noBlank.tex
\author{P.~Abreu}
\affiliation{LIP and Instituto Superior T\'{e}cnico, Technical University of Lisbon, Portugal}
\author{M.~Aglietta}
\affiliation{Istituto di Fisica dello Spazio Interplanetario (INAF), Universit\`{a} di Torino and
Sezione INFN, Torino, Italy}
\author{M.~Ahlers}
\affiliation{University of Wisconsin, Madison, WI, USA}
\author{E.J.~Ahn}
\affiliation{Fermilab, Batavia, IL, USA}
\author{I.F.M.~Albuquerque}
\affiliation{Universidade de S\~{a}o Paulo, Instituto de F\'{\i}sica, S\~{a}o Paulo, SP, Brazil}
\author{D.~Allard}
\affiliation{Laboratoire AstroParticule et Cosmologie (APC), Universit\'{e} Paris 7, CNRS-IN2P3,
 Paris, France}
\author{I.~Allekotte}
\affiliation{Centro At\'{o}mico Bariloche and Instituto Balseiro (CNEA-UNCuyo-CONICET), San
Carlos de Bariloche, Argentina}
\author{J.~Allen}
\affiliation{New York University, New York, NY, USA}
\author{P.~Allison}
\affiliation{Ohio State University, Columbus, OH, USA}
\author{A.~Almela}
\affiliation{Universidad Tecnol\'{o}gica Nacional - Facultad Regional Buenos Aires, Buenos
Aires, Argentina}
\affiliation{Instituto de Tecnolog\'{\i}as en Detecci\'{o}n y Astropart\'{\i}culas (CNEA, CONICET,
UNSAM), Buenos Aires, Argentina}
% ...
\collaboration{The Pierre Auger Collaboration}
\noaffiliation

%% file: down-going_nus_arxiv.bbl
\begin{thebibliography}{99}

\bibitem{nu_reviews}
F.~Halzen and D.~Hooper, Rep. Prog. Phys. {\bf 65}, 1025 (2002);
P.~Bhattacharjee and G.~Sigl, Phys.~Rep.~{\bf 327}, 109 (2000);
J.~K.~Becker, Phys. Rep. {\bf 458}, 173 (2008). 

\bibitem{Auger_spectrum}
J.~Abraham {\it et al.} [Pierre Auger Collaboration], Phys. Rev. Lett. \textbf{101}, 061101 (2008);
R.~U.~Abbasi {\it et al.} [HiRes], Phys. Rev. Lett. \textbf{100}, 101101 (2008).

\bibitem{nu_GZK_Berezinsky}
V.~Beresinsky and G.~Zatsepin, Phys. Lett. B \textbf{28}, 423 (1969).

\bibitem{nu_GZK_Yoshida}
S.~Yoshida and M.~Teshima, Prog. Theor. Phys. \textbf{89}, 833 (1993).

\bibitem{nu_GZK_Engel}
R.~Engel, D.~Seckel, and T.~Stanev, Phys. Rev. D \textbf{64}, 093010 (2001).

\bibitem{nu_GZK_Ahlers}
M.~Ahlers, L.~A.~Anchordoqui, M.~C.~Gonzalez-Garcia, F.~Halzen and. S.~Sarkar, Astropart. Phys. \textbf{34}, 106 (2010).

\bibitem{nu_GZK_Kotera}
K.~Kotera, D.~Allard and A.~V.~Olinto, JCAP \textbf{10}, 013 (2010).

\bibitem{nu_GZK_BerezinskyNew}
V.~Berezinsky, A.~Gazizov, M.~Kachelriess and S.~Ostapchenko, Phys. Lett. B \textbf{695}, 13 (2011).

\bibitem{IceCube}
A.~Achterberg {\it et al.} [IceCube Collaboration],
Astropart. Phys. {\bf 26}, 155 (2006).

\bibitem{Antares}
J.~A.~Aguilar {\it et al.} [ANTARES Collaboration], Phys. Lett. B \textbf{696} 16, (2011).

\bibitem{ANITA}
P.~W.~Gorham {\it et al.} [ANITA Collaboration],
Phys. Rev. Lett. {\bf 103}, 051103 (2009);
Astropart. Phys. {\bf 32}, 10 (2009).

\bibitem{PRL_nutau}
J.~Abraham {\it et al.} [Pierre Auger Collaboration],
Phys. Rev. Lett. {\bf 100}, 211101 (2008). 

\bibitem{Tiffenberg_icrc09}
J.~Tiffenberg [Pierre Auger Collaboration],
in \textit{Proceedings of the 31st International Cosmic Ray Conference, Lodz} (2009), $\#$0180, arXiv:0906.2347.

\bibitem{Gora_icrc09}
D.~Gora [Pierre Auger Collaboration],
in \textit{Proceedings of the 31st International Cosmic Ray Conference, Lodz} (2009), $\#$0077, arXiv:0906.2319.

\bibitem{nutau}  
X.~Bertou, P.~Billoir, O.~Deligny, C.~Lachaud, and A.~Letessier-Selvon, Astropart. Phys {\bf 17}, 183 (2002).

\bibitem{nu_tau_long}
J.~Abraham {\it et al.} [Pierre Auger Collaboration],
Phys. Rev. D {\bf 79}, 102001 (2009).

\bibitem{Idea_Detection}
J.~Alvarez-Mu\~niz [Pierre Auger Collaboration],
in \textit{Proceedings of the 30th International Cosmic Ray Conference, M\'erida} Vol. 4, (2007), p. 607;
I.~Vali\~no (PhD thesis) Univ. de Santiago de Compostela, ISBN: 9788497509664, (2008).

\bibitem{nu_down_Auger}
K.~S.~Capelle, J.~W.~Cronin, G.~Parente and E.~Zas, Astropart. Phys. {\bf 8}, 321 (1998); 
P.~Billoir [Pierre Auger Collaboration], J. Phys. Conf. Ser. \textbf{203}, 012125 (2010).

\bibitem{Auger_Xmax_PRL}
J.~Abraham {\it et al.} [Pierre Auger Collaboration], Phys. Rev. Lett. \textbf{104}, 091101 (2010).

\bibitem{Auger_photon_limit}
J.~Abraham {\it et al.} [Pierre Auger Collaboration], Astropart. Phys. \textbf{27}, 155 (2007);
J.~Abraham {\it et al.} [Pierre Auger Collaboration], Astropart. Phys. \textbf{29}, 243 (2008);
J.~Abraham {\it et al.} [Pierre Auger Collaboration], Astropart. Phys. {\bf 31}, 399 (2009).

\bibitem{EApaper} J.~Abraham {\it et al.} [Pierre Auger Collaboration]
Nucl. Instr. and Meth. A {\bf 523}, 50 (2004). 


\bibitem{Auger_SD}
I.~Allekotte {\it et al.} [Pierre Auger Collaboration],
Nucl. Instr. and Meth. A {\bf 586}, 409 (2008).

\bibitem{Auger_FD}
J.~Abraham {\it et al.} [Pierre Auger Collaboration], Nucl. Instr. and Meth. A \textbf{620}, 227 (2010).

\bibitem{Auger_calibration}
X.~Bertou {\it et al.} [Pierre Auger Collaboration],
Nucl. Instr. and Meth. A {\bf 568}, 839 (2006).

\bibitem{Auger_trigger}
J.~Abraham {\it et al.} 
[Pierre Auger Collaboration],  
Nucl. Instr. and Meth. A 
{\bf 613}, 29 (2010).

\bibitem{HERWIG}
G.~Corcella, I.~G.~Knowles, G.~Marchesini, S.~Moretti, K.~Odagiri, P.~Richardson, M.~H.~Seymour and B.~R.~Webber, {\sc herwig} 6.5, JHEP 0101 (2001).

\bibitem{TAUOLA}
R.~Decker, S.~Jadach, J.~H.~Kuhn and Z.~Was, Comput. Phys. Commun. {\bf 76}, 361 (1993).

\bibitem{aires}
S.~Sciutto, AIRES. Available from http://www.fisica.
unlp.edu.ar/auger/aires/.

\bibitem{Offline}
S.~Argir\`o {\it et al.} [The \Offline group - Pierre Auger Collaboration]
Nucl. Instr. and Meth. A, {\bf 580}, 1485 (2007).

\bibitem{nasa_dem} T.~G.~Farr \textit{et al.}, Rev. Geophys. \textbf{45}, 33 (2007).

\bibitem{tau_prop} O.~Blanch Bigas, O.~Deligny, K.~Payet, and V.~Van Elewyck,  Phys. Rev. D \textbf{77}, 103004 (2008).

\bibitem{thinning}
A.~M.~Hillas, in \textit{Proceedings of the 17th International
Cosmic Ray Conference, Paris} Vol. 8, (1981) p. 193.

\bibitem{Billoir_unthinning}
P.~Billoir, Astropart. Phys. {\bf 30}, 270 (2008).

\bibitem{GEANT4}
S.~Agostinelli {\it et al.}, Nucl. Instr. and Meth. A \textbf{506}, 250 (2003);
J.~Allison {\it et al.}, IEEE Trans. Nucl. Sci. \textbf{53}, 270 (2006).
See also http://geant4.web.
cern.ch/geant4/.

\bibitem{FDM}
R.~Fisher, Ann. of Eugenics {\bf7}, 179 (1936).

\bibitem{Roe_03}
B.~Roe, PHYSTAT-2003-WEJT003, 215 (2003). 

\bibitem{cooper_sarkar}
A.~Cooper-Sarkar and S.~Sarkar, JHEP {\bf 0801}, 075 (2008).

\bibitem{cteq}
J.~Pumplin, D.~R.~Stump, J.~Huston, H.~L.~Lai, P.~Nadolsky and W.~K.~Tung, JHEP \textbf{0207}, 012 (2002).

\bibitem{QGSJETII}
S.~Ostapchenko, Nucl.Phys. B - Proc. Supp. \textbf{151}, 143 (2006);
S.~Ostapchenko, Nucl.Phys. B - Proc. Supp. \textbf{151}, 147 (2006).

\bibitem{PYTHIA}
T.~Sj\"{o}strand, S.~Mrenna and P.~Skands, Comput. Phys. Commun. \textbf{178}, 11, 852 (2008).

\bibitem{Herwig++}
M.~B\"{a}hr {\it et al.}, Eur. Phys. J. C \textbf{58}, 639 (2008).

\bibitem{mstw}
A.~D.~Martin, W.~J.~Stirling, R.~S.~Thorne and G.~Watt, Eur. Phys. J. C \textbf{63}, 189 (2009).

\bibitem{corsika}
D.~Heck, J.~Knapp, J.~N.~Capdevielle, G.~Schatz and T.~Thouw, Report FZKA \textbf{6019} (1998).

\bibitem{QGSJET}
N.~Kalmykov and S.~Ostapchenko, Phys. Atom. Nucl. \textbf{56}, 346 (1993).

\bibitem{SIBYLL}
R.~Engel, T.~K.~Gaisser, T.~Stanev and P.~Lipari, in \textit{Proceedings of the 26th International Cosmic Ray Conference, Salt Lake City} Vol. 1 (1999) p. 415.

\bibitem{Connolly}
A.~Connolly, R.~S.~Thorne and D.~Waters, Phys. Rev. D \textbf{83}, 113009 (2011).

\bibitem{newSarkar}
A.~Cooper-Sarkar, P.~Mertsch and S.~Sarkar, arXiv:1106.3723, (2011).

\bibitem{Conrad_limit} J.~Conrad, O.~Botner, A.~Hallgren and C.~Perez de los Heros, Phys. Rev. D \textbf{67} (2003) 12002.
\bibitem{Feldman-Cousins} G.~J.~Feldman and R.~D.~Cousins, Phys. Rev. D \textbf{57}, 3873 (1998).

\bibitem{nu_limits}
M. Ackermann {\it et al.} [AMANDA], Astrophys. J. 675, 1014 (2008);
R. Abbasi {\it et al.} [IceCube], Phys. Rev. D {\bf 83},092003 (2011);
I. Kravchenko {\it et al.} [RICE], Phys. Rev. D {\bf 73}, 082002 (2006);
P. W. Gorham {\it et al.} [ANITA-II Collaboration], Phys. Rev. D \textbf{82} 022004 (2010), Erratum arXiv:1011.5004v1 [astro-ph];
R. Abbasi {\it et al.} [HiRes], Astrophys. J. 684, 790 (2008);
K. Martens [HiRes], arXiv:0707.4417.

\bibitem{AnchoDiffLimit} L.~A.~Anchordoqui, J.~L.~Feng, H.~Goldberg and A.~D.~Shapere, Phys. Rev. D \textbf{66}, 103002 (2002).

\bibitem{Sigl} O.~E.~Kalashev, V.~A.~Kuzmin, D.~V.~Semikoz and G.~Sigl, Phys. Rev. D \textbf{66}, 063004 (2002).

\bibitem{MPR01} K.~Mannheim, R.~J.~Protheroe and J.~P.~Rachen, Phys. Rev. D  \textbf{63} 23003, (2000).

\bibitem{BBR}  J.~K.~Becker, P.~L.~Biermann and W.~Rhode, Astropart. Phys. \textbf{23} 355, (2005).


\end{thebibliography}
